\documentclass[draftclsnofoot,10pt,onecolumn]{IEEEtran}

\usepackage[english]{babel} % Different languages:
\usepackage{amsmath}                % AMS Fonts
\usepackage{amsfonts}               % AMS Fonts 
\usepackage{amssymb}                % Mathemat. symbols of AMS
\usepackage{amsopn}                 % Further math. operators like
\allowdisplaybreaks[3] % Page breaks within eqnarray and IEEEeqnarray
\usepackage{mathrsfs}               % new mathfonts: \mathscr{ABC}
\usepackage{dsfont}
\usepackage{calc}                   % Calculations within the tex-file
\usepackage{graphicx}        % Further graphic-commands (use

\usepackage{verbatim}               % citations directly included
\usepackage{color}
\usepackage[bf]{caption}

\usepackage{cite}

\usepackage{pgfplots}
\usepackage{tikz}
\usepackage{grffile}
\pgfplotsset{compat=newest}
\usetikzlibrary{plotmarks}
\usetikzlibrary{patterns,shapes,fit}
\usepackage{amsmath}
\usepgfplotslibrary{fillbetween}

\usepackage{vmr-symbols-vecbold}
\safemath{\txant}{m\sub{t}} %number of transmit antennas
\safemath{\txantalt}{\widetilde{m}\sub{t}}
\safemath{\rxant}{m\sub{r}} %number of receive antennas
\safemath{\cohtime}{n\sub{c}} %time-frequency coherence length (in symbol times)
\safemath{\bl}{n} %blocklength
\safemath{\err}{\epsilon} %error rate
\let\snr\undefined
\safemath{\snr}{\rho} %snr
\safemath{\tfdiv}{\ell}  %amount of time-frequency diversity available (number of independent fading realizations observed over a codeword)
\safemath{\ncod}{M}
\safemath{\realset}{\amsbb{R}}
\safemath{\naturalset}{\amsbb{N}}
\safemath{\Rmax}{R^*}
\safemath{\Rmaxala}{R^*\sub{ala}}
\safemath{\Rmaxp}{\Rmax(\tfdiv,\cohtime,\epsilon,\snr)}
\safemath{\Rmaxalap}{\Rmaxala(\tfdiv,\cohtime,\epsilon,\snr)}
\safemath{\Cerg}{C\sub{erg}}
\safemath{\Cout}{C_{\text{out},\epsilon}}
\safemath{\Pout}{P\sub{out}}
\safemath{\diversity}{d}
\safemath{\multiplexing}{r}
\safemath{\Pustm}{P_{\rmatX}^{\mathrm{u}}}
\safemath{\maxantalt}{p}
\safemath{\minantalt}{q}
\safemath{\altgamma}{\widetilde{\gamma}}
\safemath{\alshouffle}{e}

\usepackage{macros}

\newtheorem{theorem}{Theorem}
\newtheorem{lemma}[theorem]{Lemma}
\newtheorem{corollary}[theorem]{Corollary}

\newtheorem{remark}{Remark}

\renewcommand{\I}[1]{\mathds{1}\left\{#1\right\}}

\title{Converse Bounds for\\ Entropy-Constrained Quantization\\ Via a Variational Entropy Inequality}

\author{Tobias Koch and Gonzalo Vazquez-Vilar}

\date{}

\begin{document}

\maketitle

\begin{abstract}
We derive a lower bound on the smallest output entropy that can be achieved via vector quantization of a $d$-dimensional source with given expected $r$th-power distortion. Specialized to the one-dimensional case, and in the limit of vanishing distortion, this lower bound converges to the output entropy achieved by a uniform quantizer, thereby recovering the result by Gish and Pierce that uniform quantizers are asymptotically optimal as the allowed distortion tends to zero. Our lower bound holds for all $d$-dimensional memoryless sources having finite differential entropy and whose integer part has finite entropy. In contrast to Gish and Pierce, we do not require any additional constraints on the continuity or decay of the source probability density function. For one-dimensional sources, the derivation of the lower bound reveals a necessary condition for a sequence of quantizers to be asymptotically optimal as the allowed distortion tends to zero. This condition implies that any sequence of asymptotically-optimal almost-regular quantizers must converge to a uniform quantizer as the allowed distortion tends to zero.
\renewcommand{\thefootnote}{}
\footnote{This work has received funding from the European Research Council (ERC) under the European Union's Horizon 2020 research and innovation programme (grant agreement number 714161), from the 7th European Union Framework Programme under Grant 333680, from the Ministerio de Econom\'ia y Competitividad of Spain under Grants TEC2013-41718-R,  RYC-2014-16332, IJCI-2015-27020,  TEC2015-69648-REDC, and TEC2016-78434-C3-3-R (AEI/FEDER, EU), and from the Comunidad de Madrid under Grant S2103/ICE-2845. The material in this paper was presented in part at the 2016 IEEE International Symposium on Information Theory, Barcelona, Spain, July 2016.

The authors are with the Signal Theory and Communications Department, Universidad Carlos III de Madrid, 28911, Legan\'es, Spain and also with the Gregorio Mara\~n\'on Health Research Institute (e-mails: \texttt{koch@tsc.uc3m.es} and \texttt{gvazquez@tsc.uc3m.es}).}
\end{abstract}
\setcounter{footnote}{0}

\section{Introduction}
\label{sec:intro}
Suppose we wish to quantize a memoryless source with an $r$th-power distortion not larger than $D$. More specifically, suppose a source produces the sequence of independent and identically distributed, $d$-dimensional, real-valued vectors $\{\vect{X}_k,\,k\in\Integers\}$ according to the distribution $P_{\vect{X}}$ and we employ a vector quantizer that produces a sequence of quantized symbols $\{\hat{\vect{X}}_k,\,k\in\Integers\}$ satisfying
\begin{equation}
\label{eq:distortion}
\varlimsup_{n\to\infty}\frac{1}{n}\sum_{k=1}^n \E{\|\vect{X}_k-\hat{\vect{X}}_k\|^r} \leq D
\end{equation}
for some norm $\|\cdot\|$ and some exponent $r>0$. (We use $\varlimsup$ to denote the \emph{limit superior} and $\varliminf$ to denote the \emph{limit inferior}.) Rate-distortion theory states that if for every blocklength $n$ and distortion constraint $D$ we quantize the sequence of source vectors $\vect{X}_1,\ldots,\vect{X}_n$ to one of $e^{nR}$ possible sequences of quantized symbols $\hat{\vect{X}}_1,\ldots,\hat{\vect{X}}_n$, then the smallest rate $R$ (in nats per source symbol) for which there exists a vector quantizer satisfying \eqref{eq:distortion} is given by \cite{shannon59}
\begin{equation}
\label{eq:R(D)}
R(D) = \inf_{P_{\hat{\vect{X}}|\vect{X}}} I(\vect{X};\hat{\vect{X}})
\end{equation}
where the infimum is over all conditional distributions of $\hat{\vect{X}}$ given $\vect{X}$ for which
\begin{equation}
\label{eq:D}
\E{\|\vect{X}-\hat{\vect{X}}\|^r}\leq D
\end{equation}
and where the expectation in \eqref{eq:D} is computed with respect to the joint distribution $P_{\vect{X}} P_{\hat{\vect{X}}|\vect{X}}$. Here and throughout this paper we omit the time indices where they are immaterial. The rate $R(D)$ as a function of $D$ is referred to as the \emph{rate-distortion function}.

While $R(D)$ characterizes the rate of the best vector quantizer that quantizes the source with $r$th-power distortion not exceeding $D$, sometimes quantizing blocks of $n$ source symbols may not be feasible, especially if $n$ is large (which is typically required to achieve \eqref{eq:R(D)}). In this case, it might be more practical to quantize each source symbol separately using a vector quantizer, defined as a (deterministic) mapping $q(\cdot)$ from the source alphabet $\set{X}$ to the (countable) reconstruction alphabet $\hat{\set{X}}$.

In this paper, we consider the symbol-wise quantization of $d$-dimensional source vectors. This setup is sufficiently general to comprise various problems of interest in high-resolution vector quantization. For example, it allows us to analyze the performance of quantization schemes that buffer $d$ consecutive symbols of a one-dimensional memoryless source and then quantize them using a $d$-dimensional vector quantizer. Furthermore, the quantization of stationary sources with memory can be studied by combining the analysis of symbol-wise, $d$-dimensional quantization with a limiting argument where $d\to\infty$.

We define the rate of the vector quantizer as the entropy of the quantized source symbol $\hat{\vect{X}}=q(\vect{X})$. Thus, the smallest rate of a symbol-wise quantizer satisfying the distortion constraint $D$ is given by
\begin{equation}
\label{eq:R_S(D)}
R_{r,d}(D) \triangleq \inf_{q(\cdot)} H\bigl(q(\vect{X})\bigr)
\end{equation}
where the infimum is over the set of quantizers $q(\cdot)$ satisfying \eqref{eq:D}. Since $\vect{X}$ determines the quantizer output $q(\vect{X})$, we have $H(q(\vect{X})|\vect{X})=0$ and the rate $R_{r,d}(D)$ can be written in the same form as \eqref{eq:R(D)} but with $P_{\hat{\vect{X}}|\vect{X}}$ replaced by $q(\cdot)$:
\begin{equation}
\label{eq:R_d(D)}
R_{r,d}(D) = \inf_{q(\cdot)} I\bigl(\vect{X};q(\vect{X})\bigr).
\end{equation}
Since $\hat{\vect{X}}=q(\vect{X})$ corresponds to a deterministic $P_{\hat{\vect{X}}|\vect{X}}$, it follows that $R_{r,s}(D)\geq R(D)$.

Any discrete memoryless source can be losslessly described by a variable-length code whose expected length is roughly the entropy of the source \cite{coverthomas91,alonorlitsky94}. Consequently, $R_{r,d}(D)$ is the smallest expected length of a vector quantization scheme that first quantizes each source symbol using a vector quantizer and then compresses the resulting sequence of quantized symbols using a lossless variable-length code.

In this paper, we focus on the asymptotic rate-distortion tradeoff in the limit as the permitted distortion tends to zero. Specifically, we study the asymptotic excess rate with respect to the rate-distortion function defined as
\begin{equation}
\label{eq:R_E}
\const{R}_{r,d} \triangleq \varliminf_{D\downarrow 0}\bigl\{R_{r,d}(D) - R(D)\bigr\}.
\end{equation}
For one-dimensional sources ($d=1$) and quadratic distortion ($r=2$), Gish and Pierce demonstrated that the excess rate is equal to \cite{gishpierce68}
\begin{equation}
\label{eq:Gish&Pierce}
\const{R}_{2,1} = \frac{1}{2}\log\frac{\pi e}{6}
\end{equation}
where $\log(\cdot)$ denotes the natural logarithm. They further showed that this excess rate can be achieved by a uniform quantizer, hence the well-known result that ``uniform quantizers are asymptotically optimal as the allowed distortion tends to zero."\footnote{The fact that, in the high-resolution case, the expected quadratic distortion of uniform scalar quantization exceeds the least distortion achievable by any quantization scheme by a factor of only $\pi e/6$ was already discovered by Koshelev in 1963. See \cite{grayneuhoff98} and references therein for more details.} For multi-dimensional sources, only bounds on $\const{R}_{r,d}$ are available. To obtain \eqref{eq:Gish&Pierce}, Gish and Pierce \cite{gishpierce68} imposed constraints on the continuity and decay of the probability density function (pdf) of $\vect{X}$. Furthermore, they merely provide an intuitive explanation of their converse result together with an outline of the proof---at the end of \cite[Appendix~II]{gishpierce68} they write ``The complete proof is surprisingly long and will not be given here."

The result \eqref{eq:Gish&Pierce} is equivalent to a result by Zador \cite{zador66}, which concerns the asymptotic excess distortion with respect to the distortion-rate function as the rate tends to infinity. Indeed, let $D_{r,d}(R)$ denote the minimum distortion achievable with a symbol-wise quantizer whose output has an entropy not exceeding $R$, i.e.,
\begin{equation}
\label{eq:Zador_min}
D_{r,d}(R) \triangleq \inf_{q(\cdot)} \E{\bigl\|\vect{X}-q(\vect{X})\bigr\|^r}
\end{equation}
where the infimum is over the set of quantizers $q(\cdot)$ satisfying $H\bigl(q(\vect{X})\bigr)\leq R$. Zador's theorem states that
\begin{equation}
\label{eq:Zador}
\lim_{R\to\infty} e^{\frac{r}{d} R} D_{r,d}(R) = b_{r,d} e^{\frac{r}{d} h(\vect{X})}
\end{equation}
where $b_{r,d}$ is a constant that only depends on $r$ and $d$ but not on the distribution of $\vect{X}$. Zador did not evaluate the constant $b_{r,d}$, but he did provide upper and lower bounds on $b_{r,d}$ that become tight for large $d$. Furthermore, for one-dimensional sources and quadratic distortion, it can be shown that $b_{2,1}=1/12$. Taking logarithms on both sides of \eqref{eq:Zador}, and replacing $R \leftrightarrow R_{r,d}(D)$ and $D_{r,d}(R) \leftrightarrow D$, we thus obtain that
\begin{equation}
\label{eq:Zador_(2,1)}
R_{2,1}(D) = h(X) + \frac{1}{2}\log\frac{1}{D} - \frac{1}{2}\log 12 + o_R(1)
\end{equation}
where $o_R(1)$ denotes error terms that vanish as $R$ tends to infinity. Furthermore, the rate-distortion function can be approximated as \cite{linkov65,linderzamir94,koch16}
\begin{equation}
\label{eq:SLB_(2,1)}
R(D) = h(X) + \frac{1}{2}\log\frac{1}{D} -\frac{1}{2} \log (2\pi e) + o_D(1)
\end{equation}
where $o_D(1)$ denotes error terms that vanish as $D$ tends to zero. Hence, the equivalence of Zador's theorem \eqref{eq:Zador} and Gish and Pierce's result \eqref{eq:Gish&Pierce} follows by applying \eqref{eq:Zador_(2,1)} and \eqref{eq:SLB_(2,1)} to \eqref{eq:R_E}.

While Zador's original proof of \eqref{eq:Zador} was flawed, a rigorous proof for quadratic distortion was given by Gray, Linder, and Li by using a Langrangian formulation of variable-rate vector quantization \cite{graylinderli02}. Their proof follows Zador's approach of 1) proving the result for sources with a uniform pdf on the unit cube; 2) extending it to piecewise constant pdfs on disjoint cubes of equal sides; 3) proving the result for a general pdf on a cube; and 4) proving the result for general pdfs. Gray \emph{et al.} do not impose any constraints on the continuity or decay of the pdf of $\vect{X}$, so their proof is more general than the proofs by Zador \cite{zador66} and by Gish and Pierce \cite{gishpierce68}.

In this paper, we derive a lower bound on $\const{R}_{r,d}$ that recovers \eqref{eq:Gish&Pierce} for one-dimensional sources and quadratic distortion. In contrast to \cite{graylinderli02}, our proof follows essentially along the lines outlined by Gish and Pierce \cite{gishpierce68}. We do not impose any constraints on the continuity or decay of the pdf of $\vect{X}$, so our proof is as general as the proof by Gray \emph{et al.}, and it is more general than the proof by Gish and Pierce.

For one-dimensional sources, the derivation of the lower bound reveals a necessary condition for a sequence of quantizers (parametrized by $D$) to achieve the asymptotic excess rate $\const{R}_{r,1}$. We apply this condition to the family of \emph{almost-regular quantizers}, which was introduced by Gy\"orgy and Linder in \cite{gyorgylinder02} and includes the uniform quantizers. Almost-regular quantizers are relevant because they achieve $D_{r,1}(R)$ when $r\geq 1$ \cite[Theorem~3]{gyorgylinder02}. Thus, for one-dimensional sources and $r$th-power distorion with $r\geq 1$, we can restrict ourselves to almost-regular quantizers without loss of optimality. The necessary condition implies that any sequence of almost-regular quantizers achieving $\const{R}_{r,1}$ must converge to a uniform quantizer as $D\to 0$. This suggests that asymptotically-optimal quantizers must essentially be uniform.

The rest of this paper is organized as follows. Section~\ref{sec:setup} introduces the problem setup and presents the main result of this paper, Theorem~\ref{thm:main}. Section~\ref{sec:outline} provides a back-of-the-envelope derivation of Theorem~\ref{thm:main} that serves as an outline for the proof. Section~\ref{sec:proof} contains the complete proof of this theorem. Section~\ref{sec:quantizers} presents a necessary condition for a sequence of quantizers to achieve the asymptotic excess rate. Section~\ref{sec:numerical} assesses the tightness of the lower bound presented in Theorem~\ref{thm:main} for multi-dimensional sources by numerically comparing it to several upper bounds achievable by lattice quantizers. Section~\ref{sec:conclusion} concludes the paper with a summary and discussion of the results.

\section{Problem Setup and Main Result}
\label{sec:setup}
We consider a $d$-dimensional, real-valued source $\vect{X}$ with support $\set{X}\subseteq \Reals^d$ whose distribution is absolutely continuous with respect to the Lebesgue measure, and we denote its pdf by $f_{\vect{X}}$. We require the source to satisfy the following two conditions:
\begin{enumerate}
\item[C1] $\vect{x}\mapsto f_{\vect{X}}(\vect{x}) \log f_{\vect{X}}(\vect{x})$ is integrable, ensuring that the differential entropy
\begin{equation}
h(\vect{X}) \triangleq - \int_{\set{X}} f_{\vect{X}}(\vect{x}) \log f_{\vect{X}}(\vect{x}) \d \vect{x}
\end{equation}
is well-defined and finite;
\item[C2] the integer part of the source $\vect{X}$ has finite entropy, i.e.,
\begin{equation}
\label{eq:inf_dim}
H(\lfloor\vect{X}\rfloor) < \infty.
\end{equation}
Here $\lfloor\vect{a}\rfloor$, $\vect{a}=(a_1,\ldots,a_d)\in\Reals^d$ denotes the element-wise floor function, i.e., $\lfloor\vect{a}\rfloor=(\lfloor a_1 \rfloor,\ldots,\lfloor a_d\rfloor)$ where $\lfloor a_{\ell}\rfloor$, $\ell=1,\ldots,d$ denotes the largest integer not larger than $a_{\ell}$. 
\end{enumerate}

Condition C2 requires that quantizing the source with a cubic lattice quantizer of unit-volume cells gives rise to a discrete random variable of finite entropy. This is necessary for the asymptotic excess rate $\const{R}_{r,d}$ to be well-defined. Indeed, as demonstrated in \cite{koch16}, if $H(\lfloor\vect{X}\rfloor)=\infty$ then the rate-distortion function $R(D)$ is infinite for any finite $D$. Since $R_{r,d}(D)\geq R(D)$, this implies that in this case $R_{r,s}(D)-R(D)$ is of the form $\infty-\infty$. Fortunately, Condition C2 is very mild. For example, by generalizing \cite[Proposition~1]{wuverdu10_2} to the vector case, it can be shown that it is satisfied if $\E{\log(1+\|\vect{X}\|)}<\infty$. This in turn is true, for example, for sources for which $\E{\|\vect{X}\|^{\alpha}}<\infty$ for some $\alpha>0$.

The quantity $H(\lfloor\vect{X}\rfloor)$ is intimately related with the \emph{R\'enyi information dimension} defined in \cite{renyi59}; see also \cite{kawabatadembo94,wuverdu10_2}. Indeed, it can be shown that a source vector has finite R\'enyi information dimension if, and only if, \eqref{eq:inf_dim} is satisfied \cite[Proposition~1]{wuverdu10_2}.

The quantizer is characterized by the (Borel measurable) function $q\colon\set{X}\to\hat{\set{X}}$ for some countable reconstruction alphabet $\hat{\set{X}}\subseteq\Reals^d$. Equivalently, we characterize $q(\cdot)$ by the quantization regions $\set{S}_i$, $i\in\Integers$ and corresponding reconstruction values $\hat{\vect{x}}_i$, $i\in\Integers$. Specifically, $\set{S}_i$, $i\in\Integers$ are disjoint (Borel measurable) subsets of $\Reals^d$ that together with the reconstruction values $\hat{\vect{x}}_i$, $i\in\Integers$ satisfy
\begin{subequations}
\begin{IEEEeqnarray}{lCl}
\bigcup_{i} \set{S}_i & = & \set{X} \\
q(\vect{x}) & = & \sum_i \hat{\vect{x}}_i \I{\vect{x}\in\set{S}_i}, \quad \text{for $\vect{x}\in\set{X}$}
\end{IEEEeqnarray}
\end{subequations}
where $\I{\cdot}$ denotes the indicator function. To simplify notation, we denote the Lebesgue measure of the quantization region $\set{S}_i$ by $\Delta_i$ and the probability of $\vect{X}$ being in $\set{S}_i$ by $p_i$.

The main result of this paper is a lower bound on the excess rate $\const{R}_{r,d}$ for general $r$ and $d$. For one-dimensional sources and quadratic distortion, it recovers the excess rate \eqref{eq:Gish&Pierce} by Gish and Pierce. However, in contrast to Gish and Pierce's result, our bound does not require any continuity or decay conditions on the behavior of the source pdf---it holds for all source vectors having a pdf, having finite differential entropy, and having finite R\'enyi information dimension.

\begin{theorem}[Main Result]
\label{thm:main}
Let the source vector $\vect{X}$ have a pdf, and assume that $h(\vect{X})$ and $H(\lfloor\vect{X}\rfloor)$ are finite. Then, the excess rate $\const{R}_{r,d}$, as defined in \eqref{eq:R_E}, is lower-bounded by
\begin{equation}
\label{eq:thm_main}
\const{R}_{r,d} \geq \frac{d}{r} \log\left(\frac{\Gamma(1+d/r)^{r/d} e}{1+d/r}\right)
\end{equation}
where $\Gamma(\cdot)$ denotes the Gamma function.
\end{theorem}
\begin{IEEEproof}
See Section~\ref{sec:proof}.
\end{IEEEproof}
In the one-dimensional case, \eqref{eq:thm_main} becomes
\begin{equation}
\label{eq:thm_main_1d}
\const{R}_{r,1} \geq \frac{1}{r} \log\left(\frac{\Gamma(1+1/r)^r e}{1+1/r}\right).
\end{equation}
As we shall see next, \eqref{eq:thm_main_1d} can be achieved by a uniform quantizer, so in the one-dimensional case the lower bound \eqref{eq:thm_main} is tight. Furthermore, for quadratic distortion, \eqref{eq:thm_main_1d} is equal to $1/2\log(\pi e/6)$, hence it recovers the excess rate obtained by Gish and Pierce.

To demonstrate the tightness of \eqref{eq:thm_main} in the one-dimensional case, and to assess the accuracy of \eqref{eq:thm_main} in higher-dimensional cases, we consider an upper bound on the excess rate that follows by restricting ourselves to the class of \emph{tessellating quantizers}. A polytope $\set{P}$ is \emph{tessellating} if there exists a partition of $\Reals^d$ consisting of translated and/or rotated copies of $\set{P}$; a \emph{tessellating quantizer}, denoted by $q_{\set{P}}\colon \set{X}\to\hat{\set{X}}$, is a quantizer whose quantization regions $\set{S}_i$ are translated and/or rotated copies of a tessellating convex polytope $\set{P}$ and the corresponding reconstruction values $\hat{\vect{x}}_i$ are the centroids of $\set{S}_i$. A special case of a tessellating quantizer is a \emph{lattice quantizer}, i.e., a quantizer whose quantization regions are the Voronoi cells of a $d$-dimensional lattice. Note that in the one-dimensional case the only convex polytope is the interval, so in this case the tessellating quantizer is the uniform quantizer. For the class of tessellating quantizers, Linder and Zeger \cite{linderzeger94} derived an asymptotic expression equivalent to \eqref{eq:Zador}.

\begin{theorem}[Linder and Zeger \protect{\cite[Theorem~1]{linderzeger94}}]\label{thm:linderzeger}
Let the source vector $\vect{X}$ have a pdf, and assume that $h(\vect{X})$ and $H(\lfloor\vect{X}\rfloor)$ are finite. Then, a tessellating quantizer $q_{\set{P}}(\cdot)$ with $r$th-power distortion $\E{\|\vect{X}-q_{\set{P}}(\vect{X})\|^r}=D$ and rate $R_{\set{P}}(D)\triangleq H\bigl(q_{\set{P}}(\vect{X})\bigr)$ satisfies
\begin{equation}
\label{eq:thm_linderzeger}
\lim_{D\downarrow 0} D e^{\frac{r}{d} R_{\set{P}}(D)} = \ell(\set{P}) e^{\frac{r}{d} h(\vect{X})}
\end{equation}
where $\ell(\set{P})$ denotes the normalized $r$-th moment of $\set{P}$, defined as
\begin{equation}
\ell(\set{P}) \triangleq \frac{\int_{\set{P}} \|\vect{x}-\hat{\vect{x}}\|^r \d \vect{x}}{V(\set{P})^{1+r/d}}
\end{equation}
and $V(\set{P})$ denotes the volume of $\set{P}$.
\end{theorem}
\begin{remark}
To be precise, \cite[Theorem~1]{linderzeger94} requires that $H\bigl(q_{\set{P_{\alpha}}}(\vect{X})\bigr)<\infty$ for some $\alpha>0$ rather than \eqref{eq:inf_dim}, i.e., $H(\lfloor\vect{X}\rfloor)<\infty$. (Here, $\set{P}_{\alpha}=\{x\in\Reals^d\colon x/\alpha\in\set{P}\}$ denotes the polytope $\set{P}$ rescaled by $\alpha$.) Nevertheless, its proof hinges on a lemma by Csisz\'ar (cf.\ \cite[Lemma~2]{linderzeger94}), which also applies if the condition $H\bigl(q_{\set{P_{\alpha}}}(\vect{X})\bigr)<\infty$ is replaced by \eqref{eq:inf_dim}. Specifically, by setting in \cite[Lemma~2]{linderzeger94} the partition $\set{B}_0=\{B_1,B_2,\ldots\}$ of $\Reals^d$ to be the set of $d$-dimensional cubes of unit-volume with the lower-most cornerpoint located at coordinates $\vect{i}\in\Integers^d$, this partition satisfies the lemma's conditions provided that \eqref{eq:inf_dim} holds.
\end{remark}

Taking logarithms on both sides of \eqref{eq:thm_linderzeger}, we obtain
\begin{equation}
\label{eq:R_P(D)}
R_{\set{P}}(D) = h(\vect{X}) + \frac{d}{r}\log\frac{1}{D} + \frac{d}{r}\log\ell(\set{P}) + o_D(1).
\end{equation}
Since a tessellating quantizer with $r$th-power distortion $D$ satisfies \eqref{eq:D}, the rate $R_{\set{P}}(D)$ upper-bounds $R_{r,d}(D)$. Furthermore, the rate-distortion function $R(D)$ can be lower-bounded as \cite{yamada80}
\begin{equation}
\label{eq:SLB}
R(D) \geq h(\vect{X}) + \frac{d}{r} \log\frac{1}{D} - \frac{d}{r}\log \left(\frac{r}{d}\bigl(V_d \Gamma(1+d/r)\bigr)^{r/d} e\right)
\end{equation}
where $V_d$ denotes the volume of the unit ball $\{\vect{x}\in\Reals^d\colon \|\vect{x}\|\leq 1\}$. The right-hand side (RHS) of \eqref{eq:SLB} is referred to as \emph{Shannon lower bound}. It has been demonstrated that its difference to $R(D)$ vanishes as $D$ tends to zero, provided that the source distribution satisfies certain conditions; see, e.g., \cite{linkov65,linderzamir94,koch16}. A finite-blocklength refinement of this bound can be found in~\cite{kostina15,kostina17_toappear}. Recently, it has been demonstrated that for sources with finite differential entropy the Shannon lower bound is asymptotically tight if, and only if, $H(\lfloor\vect{X}\rfloor)$ is finite \cite{koch16}. Thus, we have
\begin{equation}
\label{eq:R(D)_asym_2}
R(D) = h(\vect{X}) + \frac{d}{r} \log\frac{1}{D} - \frac{d}{r}\log \left(\frac{r}{d}\bigl(V_d \Gamma(1+d/r)\bigr)^{r/d} e\right)+ o_D(1)
\end{equation}
for the class of sources considered in this paper.

Combining \eqref{eq:R_P(D)} with \eqref{eq:R(D)_asym_2}, we obtain
\begin{equation}
\label{eq:really}
\lim_{D\downarrow 0} \bigl\{R_{\set{P}}(D)-R(D)\bigr\} = \frac{d}{r}\log \left(\frac{r}{d}\bigl(V_d \Gamma(1+d/r)\bigr)^{r/d}e\right) + \frac{d}{r}\log \ell(\set{P}).
\end{equation}
Recalling that $R_{r,d}(D)\leq R_{\set{P}}(D)$ for every $D$, this yields
\begin{equation}
\label{eq:RE_tessellating}
\const{R}_{r,d} \leq\frac{d}{r}\log \left(\frac{r}{d}\bigl(V_d \Gamma(1+d/r)\bigr)^{r/d}e\right)  + \inf_{\set{P}}\frac{d}{r}\log \ell(\set{P})
\end{equation}
where the infimum is over all $d$-dimensional, tessellating, convex polytopes $\set{P}$.

Using that in the one-dimensional case the only convex polytope is the interval, and noting that the interval has the normalized $r$-th moment
\begin{equation}
\ell(\set{P}) = \frac{1}{2^r(1+r)}
\end{equation}
the upper bound \eqref{eq:RE_tessellating} becomes in this case
\begin{equation}
\const{R}_{r,1} \leq \frac{1}{r} \log\left(\frac{\Gamma(1+1/r)^r e}{1+1/r}\right)
\end{equation}
which coincides with \eqref{eq:thm_main_1d}. Thus, in the one-dimensional case a tessellating quantizer (which in this case is the uniform quantizer) is asymptotically optimal.

\section{Derivation for One-Dimensional Sources and Certain Quantizers}
\label{sec:outline}
Before proving Theorem~\ref{thm:main}, we provide a simplified derivation of the lower bound \eqref{eq:thm_main} for one-dimensional sources ($d=1$) and quadratic distortion ($r=2$) that will serve as an outline for the complete proof of Theorem~\ref{thm:main} given in Section~\ref{sec:proof}. Particularized to this setting, Theorem~\ref{thm:main} becomes
\begin{equation}
\label{eq:outline_claim}
\const{R}_{2,1} \geq \frac{1}{2}\log\frac{\pi e}{6}.
\end{equation}
In our derivation we shall only consider quantizers satisfying
\begin{equation}
\label{eq:OL_assumption}
\sup_i \sup_{x\in\set{S}_i} (x-\hat{x}_i)^2 \leq \alpha D, \quad \text{for some constant $\alpha$.}
\end{equation}
This simplifying assumption is, for example, satisfied by the uniform quantizer when $\hat{x}_i$ is the midpoint of $\set{S}_i$ and the cell length $\Delta$ vanishes proportionally to $\sqrt{D}$. However, it is \emph{prima facie} unclear whether \eqref{eq:OL_assumption} holds without loss of optimality for general sources.

By \eqref{eq:R_d(D)}, we have
\begin{equation}
\label{eq:OL_R(D)_S}
R_{2,1}(D) = \inf_{q(\cdot)} I(X;\hat{X}) = h(X) - \sup_{q(\cdot)} h(X|\hat{X}).
\end{equation}
We upper-bound $h(X|\hat{X})$ by using that, conditioned on $\hat{X}=\hat{x}_i$, the support of $X$ is $\set{S}_i$, so a uniform distribution over $\set{S}_i$ maximizes the differential entropy \cite[Theorem~11.1.1]{coverthomas91}:
\begin{equation}
\label{eq:OL_uniform}
h(X|\hat{X}=\hat{x}_i) \leq \log\Delta_i.
\end{equation}
Averaging over $\hat{X}$ then yields
\begin{equation}
R_{2,1}(D) \geq h(X) - \sup_{q(\cdot)} \sum_{i} p_i \log\Delta_i.
\end{equation}
By Jensen's inequality, this can be further lower-bounded by
\begin{equation}
\label{eq:OL_1}
R_{2,1}(D) \geq h(X) - \frac{1}{2}\log\left(\sup_{q(\cdot)} \sum_i p_i \Delta_i^{2}\right).
\end{equation}
Together with \eqref{eq:SLB_(2,1)}, this yields
\begin{IEEEeqnarray}{lCl}
\varliminf_{D\downarrow 0} \bigl\{R_{2,1}(D)-R(D)\bigr\} & \geq & \varliminf_{D\downarrow 0}  \left\{\frac{1}{2}\log D + \frac{1}{2}\log(2\pi e) - \frac{1}{2} \log\left(\sup_{q(\cdot)} \sum_i p_i \Delta_i^{2}\right)\right\}. \label{eq:OL_juhu}
\end{IEEEeqnarray}
In order to prove \eqref{eq:outline_claim}, it remains to show that, for any sequence of quantizers (parametrized by $D$),
\begin{equation}
\label{eq:OL_E[L^2]}
\varlimsup_{D\downarrow 0} \frac{1}{D} \sum_{i} p_i \Delta_i^{2} \leq 12.
\end{equation}
Then the RHS of \eqref{eq:OL_juhu} is lower-bounded by $1/2 \log(\pi e/6)$ and we obtain \eqref{eq:outline_claim} upon noting that the left-hand side (LHS) of \eqref{eq:OL_juhu} is equal to $\const{R}_{2,1}$. Hence we recover Theorem~\ref{thm:main} for one-dimensional sources and quadratic distortion.

The upper bound \eqref{eq:OL_E[L^2]} follows along the lines of the proof of \cite[Lemma~1]{linderzeger94}. We first express $\E{(X-\hat{X})^2}$ as
\begin{IEEEeqnarray}{lCl}
\E{(X-\hat{X})^2} & = & \sum_{i} \int_{\set{S}_i} f_{X}(x) (x-\hat{x}_i)^2 \d x \nonumber\\
& = & \sum_i p_i \frac{1}{\Delta_i} \int_{\set{S}_i} (x-\hat{x}_i)^2 \d x - \sum_i \int_{\set{S}_i} \left[\frac{p_i}{\Delta_i} - f_{X}(x)\right] (x-\hat{x}_i)^2\d x. \label{eq:OL_2}
\end{IEEEeqnarray}
We next note that the region $\set{S}_{i}$ of measure $\Delta_i$ that minimizes $\int_{\set{S}_{i}}(x-\hat{x}_i)^2\d x$ is the interval $\bigl[\hat{x}_i-\frac{\Delta_{i}}{2},\hat{x}_i+\frac{\Delta_{i}}{2}\bigr]$, so
\begin{equation}
\frac{1}{\Delta_i} \int_{\set{S}_i} (x-\hat{x})^2\d x \geq \frac{\Delta_i^2}{12}.
\end{equation}
The first term on the RHS of \eqref{eq:OL_2} can therefore be lower-bounded by
\begin{equation}
\label{eq:OL_first}
\sum_i p_i \frac{1}{\Delta_i} \int_{\set{S}_i} (x-\hat{x}_i)^2 \d x \geq \sum_i p_i \frac{\Delta_i^{2}}{12}.
\end{equation}
To evaluate the second term on the RHS of \eqref{eq:OL_2}, we introduce the piecewise-constant pdf
\begin{equation}
\label{eq:OL_f_X^(D)}
f_{X}^{(\Delta)}(x) \triangleq \sum_{i} \frac{p_i}{\Delta_i} \I{x\in\set{S}_i}, \quad x\in\Reals.
\end{equation}
With this, we can upper-bound the second term on the RHS of \eqref{eq:OL_2} as
\begin{IEEEeqnarray}{lCl}
\sum_i \int_{\set{S}_i} \left[\frac{p_i}{\Delta_i} - f_{X}(x)\right] (x-\hat{x}_i)^2 \d x & = & \sum_i \int_{\set{S}_i} \left[f_{X}^{(\Delta)}(x) - f_{X}(x)\right] (x-\hat{x}_i)^2\d x\nonumber\\
& \leq & \alpha D \int \left|f_{X}^{(\Delta)}(x)-f_{X}(x)\right| \d x \label{eq:OL_second}
\end{IEEEeqnarray}
since, by \eqref{eq:OL_assumption}, we have $\sup_i \sup_{x\in\set{S}_i}(x-\hat{x}_i)^2\leq \alpha D$.

By Lebesgue's differentiation theorem, $f_{X}^{(\Delta)}$ converges to $f_{X}$ almost everywhere as $\sup_i \Delta_i\to 0$. It therefore follows from Scheffe's Lemma \cite[Theorem~16.12]{billingsley95} that
\begin{equation}
\label{eq:OL_scheffe}
\lim_{D\downarrow 0} \int \left|f_{X}^{(\Delta)}(x)-f_{X}(x)\right| \d x = 0.
\end{equation}
Combining \eqref{eq:OL_first} and \eqref{eq:OL_second} with \eqref{eq:OL_2}, and using that $\E{(X-\hat{X})^2}\leq D$, we obtain
\begin{equation}
\label{eq:OL_last}
 \sum_i p_i \Delta_i^{2} \leq 12 D \left(1+\alpha \int \left|f_{X}^{(\Delta)}(x)-f_{X}(x)\right| \d x\right).
\end{equation}
Together with \eqref{eq:OL_scheffe} this proves \eqref{eq:OL_E[L^2]}.

\section{Proof of Theorem~\ref{thm:main}}
\label{sec:proof}

\subsection{Variational Entropy Inequality and Auxiliary Results}

The above back-of-the-envelope derivation directly generalizes to multi-dimensional sources and $r$th-power distortion. In order to prove Theorem~\ref{thm:main}, it would remain to show that \eqref{eq:OL_assumption} holds without loss of optimality. Unfortunately, for general sources this appears to be a difficult task. Indeed, the quantization regions of the optimal quantizer are difficult to characterize since the optimal quantizer (and hence the number of quantization regions together with their locations and volumes) changes with $D$. To sidestep this problem, we replace \eqref{eq:OL_uniform} by an upper bound on $h(\vect{X}|\hat{\vect{X}}=\hat{\vect{x}}_i)$ that is based on the following variational bound on differential entropy.

\begin{lemma}
\label{lemma:duality}
Let $f$ and $g$ be arbitrary pdfs. If $- \int f(x)\log f(x)\d x$ is finite, then $-\int f(x) \log g(x) \d x$ exists and
\begin{equation}
\label{eq:lemma_duality}
- \int f(x) \log f(x) \d x \leq - \int f(x) \log g(x)\d x
\end{equation}
with equality if, and only if, $f(x)=g(x)$ almost everywhere.
\end{lemma}
\begin{IEEEproof}
See \cite[Lemma~8.3.1]{Ash90}.
\end{IEEEproof}
The inequality \eqref{eq:lemma_duality} is a direct consequence of the information inequality. Lemma~\ref{lemma:duality} is also reminiscent of \cite[Theorem~5.1]{lapidothmoser03_3}, which provides an upper bound on the mutual information between a channel input $X$ and a channel output $Y$ and holds for general random variables. In fact, when $Y$ is a real-valued random variable and the conditional distribution of $Y$ given $X$ is absolutely continuous with respect to the Lebesgue measure, then \cite[Theorem~5.1]{lapidothmoser03_3} essentially provides an upper bound on $h(Y)$ that is of the form \eqref{eq:lemma_duality}.

Lemma~\ref{lemma:duality} allows us to upper-bound differential entropy by replacing the true pdf $f$ inside the logarithm by an auxiliary pdf $g$. In order to upper-bound the conditional differential entropy $h(\vect{X}|\hat{\vect{X}}=\hat{\vect{x}}_i)$, we apply Lemma~\ref{lemma:duality} with the conditional pdf 
\begin{equation}
\label{eq:OL_g}
g_{\vect{X}|\hat{\vect{X}}}(\vect{x}|\hat{\vect{x}}_i) = \begin{cases} \frac{1}{\const{K}_{i,\eps}}, \quad & \vect{x}\in\set{B}_{i,\eps} \\ \frac{1}{\const{K}_{i,\eps}} \frac{r}{\delta^{d/r}} e^{-\frac{\|\vect{x}-\hat{\vect{x}}_i\|^r}{D\delta}}, \quad & \vect{x}\in\bar{\set{B}}_{i,\eps} \end{cases}
\end{equation}
where
\begin{subequations}
\begin{IEEEeqnarray}{lCl}
\set{B}_{i,\eps} & \triangleq & \{\vect{x}\in\set{S}_i\colon \|\vect{x}-\hat{\vect{x}}_i\| \leq \eps\}, \label{eq:Bieps}\\
\bar{\set{B}}_{i,\eps} & \triangleq & \{\vect{x}\in\set{S}_i\colon \|\vect{x}-\hat{\vect{x}}_i\| > \eps\}, \\
\const{K}_{i,\eps} & \triangleq &\Lambda_{i,\eps} + \frac{r}{\delta^{d/r}}\int_{\bar{\set{B}}_{i,\eps}} e^{-\frac{\|\vect{x}-\hat{\vect{x}}_i\|^r}{D\delta}} \d \vect{x}, \label{eq:Kie}
\end{IEEEeqnarray}
\end{subequations}
$\Lambda_{i,\eps}$ denotes the Lebesgue measure of $\set{B}_{i,\eps}$, and $\delta$ and $\eps$ are parameters to be specified later.

This conditional pdf of $\vect{X}$ given $\hat{\vect{X}}$ is uniform on a set of measure $\Lambda_{i,\eps}$ around $\hat{\vect{x}}_i$ and then decays exponentially. Intuitively, if $\eps^d$ decays more slowly than $\Delta_i$ as $D$ tends to zero, then with high probability $\vect{X}$ lies in $\set{B}_{i,\eps}$ and the upper bound obtained from Lemma~\ref{lemma:duality} is essentially equivalent to \eqref{eq:OL_uniform} but with $\Delta_i$ replaced by $\Lambda_{i,\eps}$. Our choice of $g_{\vect{X}|\hat{\vect{X}}}$ for $\vect{x}\in\bar{\set{B}}_{i,\eps}$ allows us to control the contribution of $\vect{x}$'s lying outside of $\set{B}_{i,\eps}$. We next need to show that
\begin{equation}
\label{eq:OL_toshow}
\varlimsup_{D\downarrow 0 } \frac{1}{D} \sum_i p_i \Lambda^r_{i,\eps} \leq V_d^{r/d}\left(1+\frac{r}{d}\right)
\end{equation}
which corresponds to \eqref{eq:OL_E[L^2]} generalized to arbitrary $d$ and $r$, but with $\Delta_i$ replaced by $\Lambda_{i,\eps}$. By construction of $\set{B}_{i,\eps}$, we have that $\sup_i \sup_{\vect{x}\in\set{B}_{i,\eps}} \|\vect{x}-\hat{\vect{x}}_i\|^r\leq \eps^r$, so $\set{B}_{i,\eps}$ satisfies \eqref{eq:OL_assumption} upon choosing $\eps^r = D/\kappa$ (for some constant $\kappa$). The claim \eqref{eq:OL_toshow} follows therefore immediately from the steps \eqref{eq:OL_2}--\eqref{eq:OL_last}. Thus, by using Lemma~\ref{lemma:duality} together with \eqref{eq:OL_g}, we can replace $\Delta_i$ (whose behavior as a function of $D$ is unknown) by $\Lambda_{i,\eps}$ (whose behavior can be controlled by cleverly choosing $\eps$).

Before we set out to prove Theorem~\ref{thm:main}, we first provide a number of auxiliary results that we shall need throughout the proof. The proof of Theorem~\ref{thm:main} is then given in Section~\ref{sub:main_proof}.

\begin{lemma}
\label{lemma:2}
The normalizing constant $\const{K}_{i,\eps}$ is upper-bounded by
\begin{equation}
\label{eq:lemma2}
\const{K}_{i,\eps} \leq \Lambda_{i,\eps} + d V_d D^{d/r} \Gamma\left(\frac{d}{r},\frac{\eps^r}{D\delta}\right) \leq \eps^d V_d + d V_d D^{d/r} \Gamma\left(\frac{d}{r}\right)
\end{equation}
where $\Gamma(\cdot,\cdot)$ denotes the upper incomplete Gamma function.
\end{lemma}
\begin{IEEEproof}
The first inequality in \eqref{eq:lemma2} follows from the definition of $\const{K}_{i,\eps}$ \eqref{eq:Kie} and by upper-bounding the integral on the RHS of \eqref{eq:Kie}. Indeed, since $\bar{\set{B}}_{i,\eps}\subseteq \{\vect{x}\in\Reals^d\colon \|\vect{x}-\hat{\vect{x}}_i\|>\eps\}$,
\begin{IEEEeqnarray}{lCl}
 \frac{r}{\delta^{d/r}}\int_{\bar{\set{B}}_{i,\eps}} e^{-\frac{\|\vect{x}-\hat{\vect{x}}_i\|^r}{D\delta}} \d \vect{x} & \leq & \frac{r}{\delta^{d/r}} \int_{\|\vect{x}-\hat{\vect{x}}_i\|>\eps}  e^{-\frac{\|\vect{x}-\hat{\vect{x}}_i\|^r}{D\delta}} \d \vect{x} \nonumber\\
 & = & d V_d \frac{r}{\delta^{d/r}} \int_{\rho>\eps} \rho^{d-1} e^{-\frac{\rho^r}{D\delta}} \d\rho\nonumber\\
 & = & d V_d D^{d/r} \int_{\xi>\frac{\eps^r}{D\delta}} \xi^{d/r-1} e^{-\xi} \d \xi \nonumber\\
 & = & d V_d D^{d/r} \Gamma\left(\frac{d}{r},\frac{\eps^r}{D\delta}\right)
\end{IEEEeqnarray}
where the second step follows by writing $\vect{x}-\hat{\vect{x}}_i$ in polar coordinates and by using that the surface area of the $d$-dimensional ball of radius $\rho=\|\vect{x}-\hat{\vect{x}}_i\|$ is $dV_d \rho^{d-1}$ (see, e.g., \cite[Eq.~(10)]{yamada80}), and the third step follows by the change of variable $\xi=\rho^r/(D\delta)$.

The second inequality in \eqref{eq:lemma2} follows by upper-bounding (see, e.g., \cite[Eq.~(7)]{yamada80})
\begin{equation}
\Lambda_{i,\eps}\leq \int_{\|\vect{x}-\hat{\vect{x}}\|\leq \eps}\d \vect{x}=\eps^d V_d
\end{equation}
and $\Gamma(d/r,x)\leq \Gamma(d/r)$, $x\geq 0$.
\end{IEEEproof}

\begin{lemma}
\label{lemma:3}
The set $\bar{\set{B}}_{i,\eps}$ satisfies
\begin{subequations}
\begin{IEEEeqnarray}{rCl}
\sum_i \Prob(\vect{X}\in\bar{\set{B}}_{i,\eps}) & \leq  & \frac{D}{\eps^r} \label{eq:lemma3_1a}\\
\sum_i \E{\|\vect{X}-\hat{\vect{x}}_i\|^r \I{\vect{X}\in\bar{\set{B}}_{i,\eps}}} & \leq & D \label{eq:lemma3_1b}.
\end{IEEEeqnarray}
\end{subequations}
\end{lemma}
\begin{IEEEproof}
We first prove \eqref{eq:lemma3_1a}. By the distortion constraint \eqref{eq:D}, and since $\bar{\set{B}}_{i,\eps}\subseteq \set{S}_i$ and $\|\vect{x}-\hat{\vect{x}}_i\|>\eps$ for $\vect{x}\in\bar{\set{B}}_{i,\eps}$, we have
\begin{IEEEeqnarray}{lCl}
D & \geq & \sum_i \int_{\set{S}_i} f_{\vect{X}}(\vect{x}) \|\vect{x}-\hat{\vect{x}}_i\|^r \d \vect{x} \nonumber\\
& \geq & \sum_i \int_{\bar{\set{B}}_{i,\eps}} f_{\vect{X}}(\vect{x}) \|\vect{x}-\hat{\vect{x}}_i\|^r \d \vect{x} \nonumber\\
& \geq & \sum_i \int_{\bar{\set{B}}_{i,\eps}} f_{\vect{X}}(\vect{x})\eps^r \d \vect{x}. \label{eq:lemma3}
\end{IEEEeqnarray}
Using that $\eps$ neither depends on $i$ nor on $\vect{x}$, \eqref{eq:lemma3_1a} follows by diving both sides of \eqref{eq:lemma3} by $\eps^r$.

To prove \eqref{eq:lemma3_1b} we use again the distortion constraint \eqref{eq:D} and that $\bar{\set{B}}_{i,\eps}\subseteq \set{S}_i$ to obtain
\begin{IEEEeqnarray}{lCl}
\sum_i \E{\|\vect{X}-\hat{\vect{x}}_i\|^r \I{\vect{X}\in\bar{\set{B}}_{i,\eps}}} & = & \sum_i \int_{\bar{\set{B}}_{i,\eps}} f_{\vect{X}}(\vect{x}) \|\vect{x}-\hat{\vect{x}}_i\|^r \d \vect{x} \nonumber\\
& \leq & \E{\|\vect{X}-\hat{\vect{X}}\|^r} \nonumber\\
& \leq & D.
\end{IEEEeqnarray}
\end{IEEEproof}

\subsection{Proof of Theorem~\ref{thm:main}}
\label{sub:main_proof}
Expanding $I(\vect{X};\hat{\vect{X}})$ as $h(\vect{X})-h(\vect{X}|\hat{\vect{X}})$, we obtain from \eqref{eq:R_d(D)} and \eqref{eq:R(D)_asym_2} that the excess rate can be expressed as
\begin{equation}
\label{eq:mainproof_first}
\const{R}_{r,d} = \varliminf_{D\downarrow 0} \left\{\frac{d}{r}\log D + \frac{d}{r}\log \left(\frac{r}{d}\bigl(V_d \Gamma(1+d/r)\bigr)^{r/d} e\right) - \sup_{q(\cdot)} h(\vect{X}|\hat{\vect{X}})\right\}.
\end{equation}
To derive the lower bound \eqref{eq:thm_main} given in Theorem~\ref{thm:main}, it remains to show that
\begin{equation}
\label{eq:proof_obj}
\varlimsup_{D\downarrow 0}\left\{\sup_{q(\cdot)} h(\vect{X}|\hat{\vect{X}}) -\frac{d}{r} \log D\right\} \leq \frac{d}{r}\log\bigl(V_d^{r/d}(1+r/d)\bigr).
\end{equation}
To this end, we upper-bound the conditional differential entropy $h(\vect{X}|\hat{\vect{X}})$ using Lemma~\ref{lemma:duality} together with \eqref{eq:OL_g}. This yields for every $\hat{\vect{X}}=\hat{\vect{x}}_i$
\begin{IEEEeqnarray}{lCl}
h(\vect{X}|\hat{\vect{X}}=\hat{\vect{x}}_i) & \leq & \log\const{K}_{i,\eps} -\Econd{\log\left(\frac{r}{\delta^{d/r}}e^{-\frac{\|\vect{X}-\hat{\vect{x}}_i\|^r}{D\delta}}\right)\I{\vect{X}\in\bar{\set{B}}_{i,\eps}}}{\vect{X}\in\set{S}_i} \nonumber\\
& \leq & \log\left(\Lambda_{i,\eps} + dV_d D^{d/r} \Gamma\left(\frac{d}{r},\frac{\eps^r}{D\delta}\right)\right) + \left|\log\left(\frac{r}{\delta^{d/r}}\right)\right| \Prob\bigl(\vect{X}\in\bar{\set{B}}_{i,\eps}\bigm|\vect{X}\in\set{S}_i\bigr)\nonumber\\
& & {} + \frac{1}{D\delta} \Econd{\|\vect{X}-\hat{\vect{x}}_i\|^r\I{\vect{X}\in\bar{\set{B}}_{i,\eps}}}{\vect{X}\in\set{S}_i}
\end{IEEEeqnarray}
where the second inequality follows from the bound on $\const{K}_{i,\eps}$ presented in Lemma~\ref{lemma:2} and by upper-bounding $-\log(r/\delta^{d/r})\leq \bigl|\log(r/\delta^{d/r})\bigr|$. Averaging over $\hat{\vect{X}}$ then yields
\begin{IEEEeqnarray}{lCl}
h(\vect{X}|\hat{\vect{X}}) & \leq & \sum_{i} p_i \log\left(\Lambda_{i,\eps} + dV_d D^{d/r} \Gamma\left(\frac{d}{r},\frac{\eps^r}{D\delta}\right)\right) \nonumber\\
& & {} + \left|\log\left(\frac{r}{\delta^{d/r}}\right)\right| \sum_i \Prob\bigl(\vect{X}\in\bar{\set{B}}_{i,\eps}\bigr) + \frac{1}{D\delta} \sum_i \E{\|\vect{X}-\hat{\vect{x}}_i\|^r\I{\vect{X}\in\bar{\set{B}}_{i,\eps}}}. \label{eq:proof_1}
\end{IEEEeqnarray}
By Lemma~\ref{lemma:3}, this can be further upper-bounded by
\begin{equation}
\label{eq:proof_2}
h(\vect{X}|\hat{\vect{X}}) \leq \sum_{i} p_i \log\left(\Lambda_{i,\eps} + dV_d D^{d/r} \Gamma\left(\frac{d}{r},\frac{\eps^r}{D\delta}\right)\right) + \left|\log\frac{r}{\delta^{d/r}}\right| \frac{D}{\eps^r} + \frac{1}{\delta}.
\end{equation}
We next choose
\begin{equation}
\label{eq:proof_eps}
\eps^r = \frac{D}{\kappa}
\end{equation}
for some $\kappa>0$ that we will let tend to zero at the end of the proof. For ease of exposition, we do not always make this choice explicit in the notation but write $\eps^r$ or $D/\kappa$ depending on which is more convenient.

With this choice, the second term on the RHS of \eqref{eq:proof_2} becomes $\kappa \left|\log(r/\delta^{d/r})\right|$. To evaluate the first term on the RHS of \eqref{eq:proof_2}, we express $p_i$ as
\begin{equation}
p_i  = \Prob(\vect{X}\in\set{B}_{i,\eps}) + \Prob(\vect{X}\in\bar{\set{B}}_{i,\eps})
\end{equation}
and define
\begin{equation}
\label{eq:wpeps}
\wp_{\eps} \triangleq \sum_i \Prob(\vect{X}\in\bar{\set{B}}_{i,\eps}).
\end{equation}
By Lemma~\ref{lemma:3}, we have
\begin{equation}
\label{eq:proof_3}
\wp_{\eps} \leq \kappa
\end{equation}
which vanishes as we let $\kappa$ tend to zero. With the above definition, and applying the second inequality in \eqref{eq:lemma2} (Lemma~\ref{lemma:2}), we obtain for the first term on the RHS of \eqref{eq:proof_2} that
\begin{IEEEeqnarray}{lCl}
\IEEEeqnarraymulticol{3}{l}{\sum_{i} p_i \log\left(\Lambda_{i,\eps} + dV_d D^{d/r} \Gamma\left(\frac{d}{r},\frac{\eps^r}{D\delta}\right)\right)} \nonumber\\
\quad & = & \sum_{i} \Prob(\vect{X}\in\set{B}_{i,\eps}) \log\left(\Lambda_{i,\eps} + d V_d D^{d/r} \Gamma\left(\frac{d}{r},\frac{1}{\kappa\delta}\right)\right)  + \sum_{i} \Prob(\vect{X}\in\bar{\set{B}}_{i,\eps}) \log\left(\Lambda_{i,\eps} + d V_d D^{d/r} \Gamma\left(\frac{d}{r},\frac{1}{\kappa\delta}\right)\right) \nonumber\\
& \leq & \sum_{i} \Prob(\vect{X}\in\set{B}_{i,\eps}) \log\left(\Lambda_{i,\eps} + dV_d D^{d/r} \Gamma\left(\frac{d}{r},\frac{1}{\kappa\delta}\right)\right)  + \wp_{\eps} \log\left(V_d\frac{D^{d/r}}{\kappa^{d/r}}+dV_d D^{d/r} \Gamma(d/r)\right). \IEEEeqnarraynumspace \label{eq:proof_4}
\end{IEEEeqnarray}
Using \eqref{eq:proof_3} and that $\sum_{i}\Prob(\vect{X}\in\set{B}_{i,\eps}) + \wp_{\eps} = 1$, \eqref{eq:proof_4} becomes\begin{IEEEeqnarray}{lCl}
\IEEEeqnarraymulticol{3}{l}{\sum_{i} p_i \log\left(\Lambda_{i,\eps} + dV_d D^{d/r} \Gamma\left(\frac{d}{r},\frac{1}{\kappa\delta}\right)\right)} \nonumber\\
\quad & \leq & \sum_{i} \Prob(\vect{X}\in\set{B}_{i,\eps}) \log\left(\frac{\Lambda_{i,\eps}}{D^{d/r}} + d V_d \Gamma\left(\frac{d}{r},\frac{1}{\kappa\delta}\right)\right) + \wp_{\eps} \log\left(\frac{V_d}{\kappa^{d/r}}+dV_d\Gamma(d/r)\right) + \frac{d}{r} \log D \nonumber\\
& \leq & \frac{d}{r} \sum_{i} \Prob(\vect{X}\in\set{B}_{i,\eps})\log\left(\frac{\Lambda_{i,\eps}}{D^{d/r}} + dV_d \Gamma\left(\frac{d}{r},\frac{1}{\kappa\delta}\right)\right)^{r/d} + \kappa \log\left(\frac{V_d}{\kappa^{d/r}}+d V_d\Gamma(d/r)\right) + \frac{d}{r} \log D. \label{eq:proof_5}\IEEEeqnarraynumspace
\end{IEEEeqnarray}
By Jensen's inequality, the first term on the RHS of \eqref{eq:proof_5} is upper-bounded by
\begin{IEEEeqnarray}{lCl}
\IEEEeqnarraymulticol{3}{l}{\frac{d}{r} \sum_{i} \Prob(\vect{X}\in\set{B}_{i,\eps})\log\left(\frac{\Lambda_{i,\eps}}{D^{d/r}} + d V_d \Gamma\left(\frac{d}{r},\frac{1}{\kappa\delta}\right)\right)^{r/d}}\nonumber\\
\qquad & \leq & (1-\wp_{\eps})\frac{d}{r}\log\left(\frac{1}{1-\wp_{\eps}} \sum_{i} \Prob(\vect{X}\in\set{B}_{i,\eps})\left[\frac{\Lambda_{i,\eps}}{D^{d/r}} + d V_d \Gamma\left(\frac{d}{r},\frac{1}{\kappa\delta}\right)\right]^{r/d}\right).\label{eq:proof_6}
\end{IEEEeqnarray}
For $r/d<1$, we have $(x+\alpha)^{r/d}\leq x^{r/d} + \alpha^{r/d}$ for every $x,\alpha\geq 0$; for $r/d\geq 1$, the function $x \mapsto \left(x^{d/r} + \alpha\right)^{r/d}$ is concave for every $\alpha\geq 0$. Consequently,
\begin{IEEEeqnarray}{lCl}
\IEEEeqnarraymulticol{3}{l}{ \frac{1}{1-\wp_{\eps}}\sum_{i} \Prob(\vect{X}\in\set{B}_{i,\eps})\left[\frac{\Lambda_{i,\eps}}{D^{d/r}} + dV_d \Gamma\left(\frac{d}{r},\frac{1}{\kappa\delta}\right)\right]^{r/d}} \nonumber\\
\qquad\qquad\qquad & \leq & \begin{cases}  \frac{1}{1-\wp_{\eps}}\sum_{i} \Prob(\vect{X}\in\set{B}_{i,\eps}) \frac{\Lambda_{i,\eps}^{r/d}}{D} + d^{r/d} V_d^{r/d}  \Gamma\left(\frac{d}{r},\frac{1}{\kappa\delta}\right)^{r/d}, \quad & r/d<1 \\[5pt] \left[\left(\frac{1}{1-\wp_{\eps}}\sum_{i} \Prob(\vect{X}\in\set{B}_{i,\eps}) \frac{\Lambda_{i,\eps}^{r/d}}{D}\right)^{d/r}+dV_d\Gamma\left(\frac{d}{r},\frac{1}{\kappa\delta}\right)\right]^{r/d}, \quad & r/d \geq 1\end{cases} \label{eq:proof_6.5}
\end{IEEEeqnarray}
where the upper bound for $r/d\geq 1$ follows from Jensen's inequality.

We next generalize \eqref{eq:OL_E[L^2]}, namely,
\begin{equation}
\varlimsup_{D\downarrow 0} \frac{1}{D} \sum_{i} p_i \Delta_i^{2} \leq 12
\end{equation}
to the $d$-dimensional sets $\set{B}_{i,\eps}$ of Lebesgue measure $\Lambda_{i,\eps}$. To this end, we follow essentially the steps \eqref{eq:OL_2}--\eqref{eq:OL_last} in Section~\ref{sec:outline} with $\set{S}_i$ replaced by $\set{B}_{i,\eps}$ and with $\Delta_i$ replaced by $\Lambda_{i,\eps}$. However, \eqref{eq:OL_scheffe} is based on Lebesgue's differentiation theorem, which requires that the families of sets $\set{B}_{i,\eps}$ (parametrized by $D$) have \emph{bounded eccentricity}.\footnote{A family $\set{F}$ of sets is said to have bounded eccentricity if there exists a constant $c>0$ such that for every $\set{S}\in\set{F}$ the Lebesgue measure of $\set{S}$ is not smaller than $c$ times the volume of the smallest ball containing $\set{S}$.} Since $\set{B}_{i,\eps}$ is the intersection of $\set{S}_i$ with the $d$-dimensional ball of radius $\eps$ centered at $\hat{\vect{x}}_i$, cf.\ \eqref{eq:Bieps}, and since $\set{S}_i$ is arbitrary, the sets $\set{B}_{i,\eps}$ may not fulfill this condition. In the one-dimensional case, a sufficient condition for $\set{B}_{i,\eps}$  having bounded eccentricity would be that, for every distortion $D$, the quantization regions $\set{S}_i$ are convex. This in turn can be assumed without loss of optimality, e.g., for quadratic distortion and sources with well-behaved pdfs \cite{gyorgylinder02}. However, for one-dimensional sources with general pdfs, or for higher-dimensional sources, assuming convex quantization regions may be too restrictive. Fortunately, the families of sets $\set{B}_{i,\eps}$ that have not bounded eccentricity can be disregarded without affecting the final result. The inequality \eqref{eq:OL_E[L^2]} can therefore be generalized to the case at hand without imposing any additional constraints on the quantization regions $\set{S}_i$, $i\in\Integers$ or the source pdf $f_{\vect{X}}$.
The result is stated in the following lemma.

\begin{lemma}
\label{lemma:4}
Let the sets $\set{B}_{i,\eps}$, $i\in\Integers$ be defined in \eqref{eq:Bieps}, and let $\Lambda_{i,\eps}$, $i\in\Integers$ denote the Lebesgue measures of these sets. Assume that $\eps^r=D/\kappa$. Then, for every $\kappa>0$,
\begin{equation}
\label{eq:proof_8}
\varlimsup_{D\downarrow0} \sup_{q(\cdot)}\sum_{i} \Prob(\vect{X}\in\set{B}_{i,\eps}) \frac{\Lambda^{r/d}_{i,\eps}}{D} \leq V_d^{r/d}\left(1+\frac{r}{d}\right).
\end{equation}
\end{lemma}
\begin{IEEEproof}
See Appendix~\ref{app:lemma4}.
\end{IEEEproof}
Combining Lemma~\ref{lemma:4} with \eqref{eq:proof_2}--\eqref{eq:proof_6.5}, and bounding $0\leq \wp_{\eps} \leq \kappa$, we obtain that
\begin{subequations}
\begin{IEEEeqnarray}{lCl}
\varlimsup_{D\downarrow 0} \left\{\sup_{q(\cdot)} h(\vect{X}|\hat{\vect{X}}) -\frac{d}{r} \log D\right\} & \leq & \frac{d}{r} \log\left(\frac{V_d^{r/d}(1+r/d)}{1-\kappa} + d^{r/d}V_d^{r/d} \Gamma\left(\frac{d}{r},\frac{1}{\kappa\delta}\right)^{r/d}\right) \nonumber\\
\IEEEeqnarraymulticol{3}{r}{ {} + \kappa\log\left(\frac{V_d}{\kappa^{d/r}}+dV_d\Gamma(d/r)\right) + \kappa\left|\log\frac{r}{\delta^{d/r}}\right| + \frac{1}{\delta}, \quad \textnormal{for $r/d<1$}}\label{eq:proof_7}
\end{IEEEeqnarray}
and
\begin{IEEEeqnarray}{lCl}
\varlimsup_{D\downarrow 0} \left\{\sup_{q(\cdot)} h(\vect{X}|\hat{\vect{X}}) -\frac{d}{r} \log D\right\} & \leq & \log\left(\frac{V_d(1+r/d)^{d/r}}{(1-\kappa)^{d/r}}+dV_d \Gamma\left(\frac{d}{r},\frac{1}{\kappa\delta}\right)\right) \nonumber\\
\IEEEeqnarraymulticol{3}{r}{{} + \kappa \log\left(\frac{V_d}{\kappa^{d/r}}+dV_d\Gamma(d/r)\right) + \kappa\left|\log\frac{r}{\delta^{d/r}}\right| + \frac{1}{\delta}, \quad \textnormal{for $r/d\geq 1$.}} \label{eq:proof_7.5}
\end{IEEEeqnarray}
\end{subequations}
Using that $\lim_{\xi\to\infty} \Gamma(d/r,\xi)=0$ and $\lim_{\xi\to 0} \xi\log(\alpha/\xi^{d/r}+\beta)=0$ (for any $
\alpha,\beta>0$), letting $\kappa\to 0$ yields
\begin{equation}
\label{eq:proof_juhuu}
\varlimsup_{D\downarrow 0} \left\{\sup_{q(\cdot)} h(\vect{X}|\hat{\vect{X}}) -\frac{d}{r} \log D\right\} \leq \frac{d}{r}\log\bigl(V_d^{r/d}(1+r/d)\bigr) + \frac{1}{\delta}.
\end{equation}
This in turn proves \eqref{eq:proof_obj} upon letting $\delta\to\infty$ and concludes the proof of Theorem~\ref{thm:main}.

\section{Asymptotically Optimal Quantizers}
\label{sec:quantizers}
As mentioned at the end of Section~\ref{sec:setup}, in the one-dimensional case uniform quantizers with cells of length $2(1+r)^{1/r}D^{1/r}$ achieve the asymptotic excess rate $\const{R}_{r,1}$. Hence, uniform quantizers are asymptotically optimal as the allowed distortion tends to zero. One may wonder whether every sequence of quantizers achieving $\const{R}_{r,1}$ must converge to a uniform quantizer as $D\to 0$, or whether uniform quantizers are merely a convenient choice and other quantizers with vanishing cells are also asymptotically optimal. In this section, we partially address this question by presenting in Theorem~\ref{thm:necessary} a necessary condition for the asymptotic optimality of a sequence of quantizers (parametrized by $D$). We then apply this condition to the family of almost-regular quantizers.
\begin{theorem}
\label{thm:necessary}
Suppose the sequence of quantizers $q(\cdot)$ (parametrized by $D$) with quantization regions $\set{S}_i$, $i\in\Integers$ satisfying the distortion constraint $\E{|X-q(X)|^r}\leq D$ achieves the asymptotic excess distortion
\begin{equation}
\varliminf_{D\downarrow 0} \left\{H\bigl(q(X)\bigr)-R(D)\right\} = \frac{1}{r} \log\left(\frac{\Gamma(1+1/r)^r e}{1+1/r}\right).
\end{equation}
Then,
\begin{equation}
\label{eq:thm_necessary}
\lim_{\rho\to\infty} \varlimsup_{D\downarrow 0} \sum_{i} \Prob(X\in\set{S}_i) \I{\left|\frac{\Lambda_{i,\rho D^{1/r}}^r}{D}- 2^r(1+r)\right|\leq\vartheta} = 1, \quad \text{for every $\vartheta>0$.}
\end{equation}
Here, $\Lambda_{i,\rho D^{1/r}}$ denotes the Lebesgue measure of $\set{B}_{i,\eps}$ in \eqref{eq:Bieps} for $\eps=\rho D^{1/r}$.
\end{theorem}
\begin{IEEEproof}
This result is a direct consequence of Jensen's inequality applied in \eqref{eq:proof_6} in the proof of Theorem~\ref{thm:main}. See Appendix~\ref{app:thmnecessary} for a detailed proof.
\end{IEEEproof}

If we interpret the quantizer as a random variable that takes on the value $\set{S}_i$ with probability $\Prob(X\in\set{S}_i)$, then Theorem~\ref{thm:necessary} can be paraphrased as follows: ``A sequence of quantizer achieves the asymptotic excess distortion $\const{R}_{r,1}$ only if $\Lambda_{i,\rho D^{1/r}}$ converges in probability to $2(1+r)^{1/r} D^{1/r}$ as $D\to 0$ and $\rho\to\infty$."

A quantizer $q(\cdot)$ is said to be \emph{almost regular} if there exists a set $\bar{\set{S}}\subset \set{X}$ of Lebesgue measure zero such that on $\set{X}\setminus\bar{\set{S}}$ the quantization regions are intervals containing   the reconstruction value \cite{gyorgylinder02}. (For all $x\in\bar{\set{S}}$, we can define $q(x)$ in an arbitrary manner without changing the entropy and distortion of $q(\cdot)$.) In other words, an almost-regular quantizer $q(\cdot)$ can be written as
\begin{subequations}
\begin{IEEEeqnarray}{lCll}
q(x) & = & \sum_i c_i \I{a_i \leq x < b_i}, \quad & \text{for $x\in\set{X}\setminus\bar{\set{S}}$} \\
q(x) & = & \sum_i \bar{x}_i \I{x\in\bar{\set{S}}_i}, \quad & \text{for $x\in\bar{\set{S}}$}
\end{IEEEeqnarray}
\end{subequations}
where $a_i\leq c_i\leq b_i$, and where $\bar{x}_i$ and $\bar{\set{S}}_i$ are arbitrary.

For almost-regular quantizers, condition \eqref{eq:thm_necessary} in Theorem~\ref{thm:necessary} can be simplified as follows. Firstly, since the source has a pdf and $\bar{\set{S}}$ has measure zero,
\begin{equation}
\sum_i \Prob(X\in\set{S}_i\cap \bar{\set{S}}) = 0.
\end{equation}
Secondly, for any quantization region $[a_i,b_i)\subseteq \set{X}\setminus\bar{\set{S}}$ and reconstruction value $c_i\in[a_1,b_i)$, we have
\begin{equation}
\label{eq:arq_1}
\min\left\{\Delta^r_i,\rho^r D\right\} \leq \Lambda^r_{i,\rho D^{1/r}} \leq \Delta^r_i
\end{equation}
where $\Delta_i=b_i-a_i$. Consequently,
\begin{equation}
\I{\left|\frac{\Lambda_{i,\rho D^{1/r}}^r}{D}- 2^r(1+r)\right|\leq\vartheta} = \I{\left|\frac{\Delta_{i}}{D}- 2^r(1+r)\right|\leq\vartheta}, \qquad \rho \geq \left(2^r(1+r)\right)^{1/r}.
\end{equation}
We thus have the following result:

\begin{corollary}
\label{cor}
Suppose the sequence of almost-regular quantizers $q(\cdot)$ (parametrized by $D$) with quantization regions $\set{S}_i$, $i\in\Integers$ satisfying the distortion constraint $\E{|X-q(X)|^r}\leq D$ achieves the asymptotic excess distortion
\begin{equation}
\varliminf_{D\downarrow 0} \left\{H\bigl(q(X)\bigr)-R(D)\right\} = \frac{1}{r} \log\left(\frac{\Gamma(1+1/r)^r e}{1+1/r}\right).
\end{equation}
Then,
\begin{equation}
\varlimsup_{D\downarrow 0} \sum_i \Prob(X\in\set{S}_i) \I{\left|\frac{\Delta_{i}}{D}- 2^r(1+r)\right|\leq\vartheta} = 1, \qquad \text{for every $\vartheta>0$.}
\end{equation}
Here, $\Delta_i$ denotes the Lebesgue measure of $\set{S}_i$.
\end{corollary}

Again, interpreting the quantizer as a random variable that takes on the value $\set{S}_i$ with probability $\Prob(X\in\set{S}_i)$, Corollary~\ref{cor} can be paraphrased as ``any sequence of almost-regular quantizers achieving $\const{R}_{r,1}$ must converge in probability to a uniform quantizer as $D\to 0$."

\section{Balls versus Tessellating Polytopes}
\label{sec:numerical}
The lower bound \eqref{eq:thm_main} on the excess rate presented in Theorem~\ref{thm:main} hinges on the fact that the distortion over the quantization region $\set{S}_i$, i.e., $\int_{\set{S}_i} \|\vect{x}-\hat{\vect{x}}\|^r \d\vect{x}$, is lower-bounded by the distortion over a ball around $\hat{\vect{x}}_i$ with the same volume (cf.~\eqref{eq:proof_ULB} in the proof of Theorem~\ref{thm:main} with $\set{B}_{i,\eps}$ replaced by $\set{S}_i$ and with $\Lambda_{i,\eps}$ replaced by $\Delta_i$). Since the one-dimensional ball is an interval and, hence, tessellates $\Reals$, it follows that for scalar sources the lower bound \eqref{eq:thm_main} is achieved by a tessellating quantizer, so in this case it is tight. However, it is expected that this is no longer true for multi-dimensional sources, since in general balls do not tessellate the space. In fact, it is unclear whether there exists any (possibly non-tessellating) vector quantizer that achieves \eqref{eq:thm_main} for multi-dimensional sources.

To assess the tightness of the obtained lower bound, we compare it numerically with the excess rates achievable by several lattice quantizers. To this end, we use Linder and Zeger's upper bound for tessellating quantizers \eqref{eq:RE_tessellating} together with the normalized second moments $\ell(\set{P})$ of various lattice quantizers tabulated in \cite[Table~I]{conway85}. In order to better compare our results with previous works, in this section we consider the \emph{excess rate per dimension}, defined as $\bar{\const{R}}_{r,d}\triangleq \const{R}_{r,d}/d$. The excess rate per dimension is relevant, for example, in the analysis of quantization schemes that buffer $d$ consecutive symbols of a one-dimensional memoryless source and then quantize them using a $d$-dimensional vector quantizer.

For the sake of simplicity, we only consider quadratic distortion and the Euclidean norm. In this case, the lower bound \eqref{eq:thm_main} becomes
\begin{equation}
\label{eq:num_LB}
\bar{\const{R}}_{2,d} \geq \frac{1}{2}\log\left(2\pi e\frac{\Gamma(1+d/2)^{2/d}}{\pi(2+d)}\right).
\end{equation}
Furthermore, the upper bound corresponding to tessellating quantizers \eqref{eq:RE_tessellating} becomes
\begin{equation}
\label{eq:num_UB}
\bar{\const{R}}_{2,d} \leq \frac{1}{2} \log\left(2\pi e\frac{1}{d}\inf_{\set{P}}\ell(\set{P})\right).
\end{equation}
Another upper bound on $\bar{\const{R}}_{2,d}$ follows from an upper bound on $b_{r,d}$ in Zador's theorem \eqref{eq:Zador} that was presented in \cite{zador66}. This upper bound is based on random coding arguments and yields for quadratic distortion and the Euclidean norm
\begin{equation}
\label{eq:num_zadorUB}
\bar{\const{R}}_{2,d} \leq \frac{1}{2} \log\left(2\pi e\frac{\Gamma(1+2/d)\Gamma(1+d/2)^{2/d}}{\pi d}\right).
\end{equation}
The bound \eqref{eq:num_zadorUB} demonstrates that $\bar{\const{R}}_{2,d}$ vanishes as $d$ tends to infinity. This is perhaps not very surprising, since the rate-distortion function $R(D)$ is essentially achieved by a vector quantizer whose dimension tends to infinity. 

\begin{figure}[t]
  \begin{center}
 \includegraphics{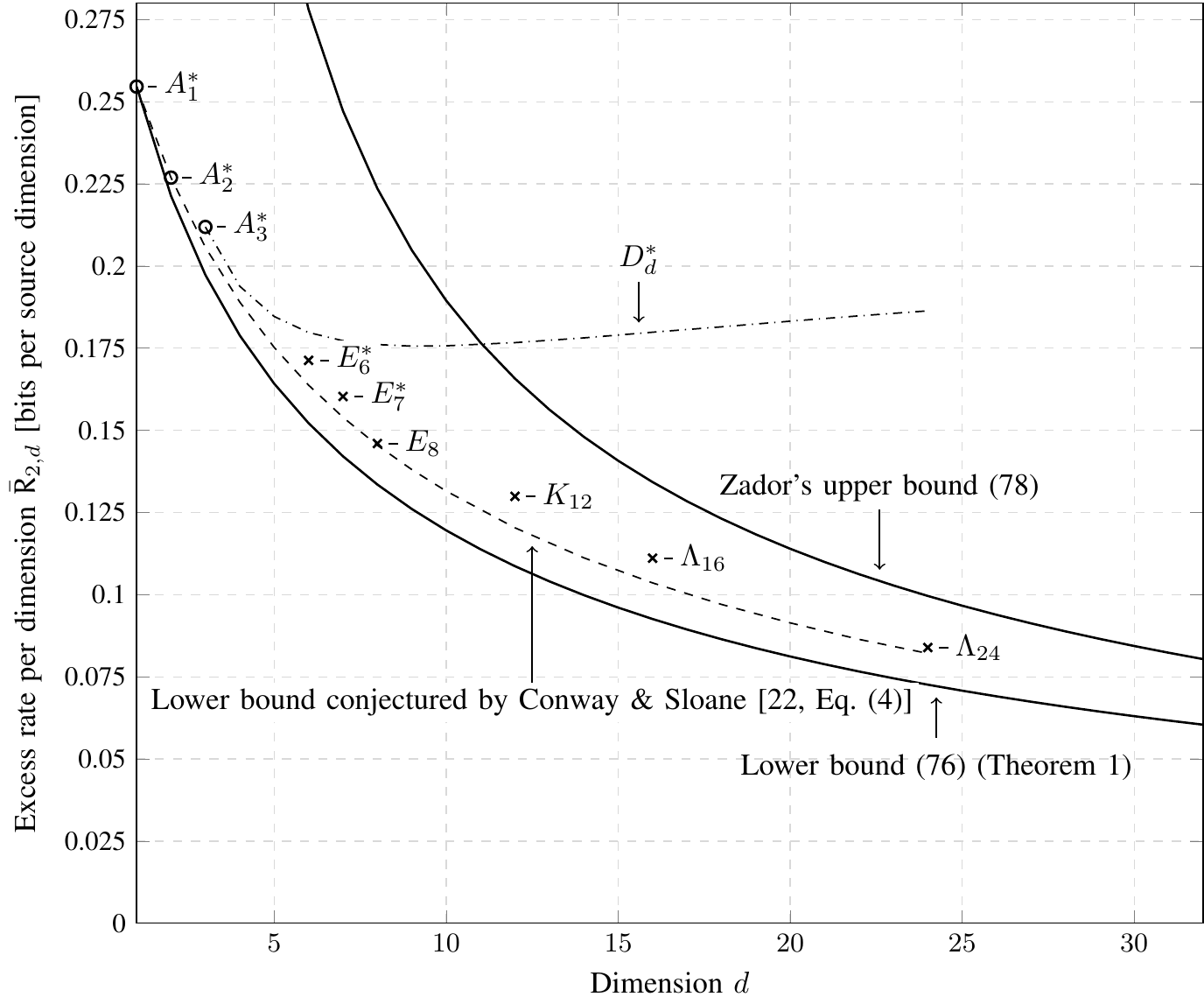}
  \caption{Bounds on the excess rate per dimension $\bar{\const{R}}_{2,d}$ (in bits per source dimension) of a $d$-dimensional vector quantizer. The excess rate per dimension attained by lattice quantizers was obtained by applying to \eqref{eq:num_UB} the normalized second moments tabulated in \cite[Table~I]{conway85}.}
	\label{fig:vector-bounds}
  \end{center}
  \vspace{-6mm}
\end{figure}

In Figure~\ref{fig:vector-bounds}, we depict the bounds \eqref{eq:num_LB} and \eqref{eq:num_zadorUB} as a function of the dimension $d$. We further show several achievability results based on lattice quantizers \eqref{eq:num_UB}. The normalized second moments $\ell(\set{P})$ corresponding to these lattice quantizers were tabulated by Conway and Sloane in \cite[Table~I]{conway85}. In fact, Figure~\ref{fig:vector-bounds} is equivalent to \cite[Figure~1]{conway85} with the only difference that here we plot the excess rate per dimension whereas Conway and Sloane plot the normalized second moment. Specifically, we include the excess rates per dimension incurred by a (one-dimensional) uniform quantizer, by a (two-dimensional) hexagonal quantizer, and by the three-dimensional tessellating quantizer whose regions are cuboctahedrons. These quantizers correspond to the so-called Voronoi lattices of the first type $A_1^*$ (the integers), $A_2^*$ (the two-dimensional hexagonal lattice), and $A_3^*$ (the body-centered cubic lattice). For $d\geq 3$, we further include the excess rates per dimension attained by the $D_d^*$ lattices. Labeled with cross markers, we show the excess rates per dimension corresponding to the lattices $E_6^*$, $E_7^*$, the Gosset lattice $E_8$, the Coxeter-Todd lattice $K_{12}$, the Barnes-Wall lattice $\Lambda_{16}$, and the Leech lattice $\Lambda_{24}$. We refer to \cite{conway85} and references therein for further details.

Finally, we compare the obtained bounds with a conjectured lower bound by Conway and Sloane \cite[Eq.~(4)]{conway85} that follows by computing the distortion attained by a set of reconstruction points located at the vertices of a $d$-dimensional tetrahedron. Note that this bound was computed for \emph{fixed-rate quantizers}, i.e., for quantizers that have a finite number $M$ of quantization regions and whose rate is defined as $\log M$. While the excess rate achievable by a fixed-rate quantizer can also be achieved by an entropy-constrained quantizer, the converse is not necessarily true. It is thus \emph{prima facie} unclear whether Conway and Sloane's conjectured lower bound would also apply to entropy-constrained quantizers. Nevertheless, we decided to include it here since it is remarkably close to the excess rates per dimension corresponding to lattices $E_8$ and $\Lambda_{24}$.

As mentioned above, the excess rate per dimension vanishes as $d$ tends to infinity. However, as illustrated by Figure~\ref{fig:vector-bounds}, it decays slowly: for example, for a 10-dimensional vector quantizer we still have
\begin{equation}
\bar{\const{R}}_{2,10} \geq \frac{1}{2}\log_2\left(\frac{e\Gamma(6)^{1/5}}{6}\right) \approx 0.1196 \text{ bits per source dimension}
\end{equation}
which is, arguably, not much smaller than the excess rate per dimension of the (one-dimensional) uniform quantizer
\begin{equation*}
\bar{\const{R}}_{2,1}=\frac{1}{2}\log_2\bigl(\pi e/6\bigr)\approx 0.2546 \text{ bits per source dimension.}
\end{equation*}
(Here $\log_2(\cdot)$ denotes the binary logarithm.) In general, the bounds on $\bar{\const{R}}_{2,d}$ given in \eqref{eq:num_LB} and \eqref{eq:num_zadorUB} are of the order $\Theta(\log d/d)$.

Observe that for multi-dimensional sources the gap between the lower bound \eqref{eq:num_LB} and the excess rate per dimension achievable with lattice quantizers is substantial. This gap is partly due to the fact that, in order to derive the lower bound \eqref{eq:thm_main}, we lower-bounded the distortion over the quantization region $\set{S}_i$ by that over a ball with the same volume, cf.~\eqref{eq:proof_ULB}. To obtain a tighter lower bound, we may need a more accurate approximation of this distortion that, like the conjectured bound by Conway and Sloane, takes the geometry of the optimal quantization regions into account. 

\section{Conclusions}
\label{sec:conclusion}
The nonnegativity of relative entropy implies that the differential entropy of a random variable $X$ with pdf $f$ is upper-bounded by $-\E{\log g(X)}$ for any arbitrary pdf $g$. Using this inequality with a cleverly chosen $g$, we derived a lower bound on the asymptotic excess rate of entropy-constrained scalar quantization. Specialized to the one-dimensional case and quadratic distortion, this bound coincides with the excess rate obtained by Gish and Pierce in \cite{gishpierce68}, and by Gray \emph{et al.} in \cite{graylinderli02} particularized for scalar quantizers. The proposed derivation thus recovers the well-known result that uniform quantizers are asymptotically optimal as the allowed distortion vanishes.

Our result holds for any $d$-dimensional memoryless source $\vect{X}$ that satisfies $|h(\vect{X})|<\infty$ and $H(\lfloor \vect{X}\rfloor)<\infty$.  The presented proof is thus as general as the proof by Gray \emph{et al.}, and it is more general than the proof by Gish and Pierce. In fact, it has recently been shown that these conditions are necessary and sufficient for the Shannon lower bound to be asymptotically tight for vanishing distortion, and that $H(\lfloor\vect{X}\rfloor)<\infty$ is a necessary and sufficient condition for the rate-distortion function to be finite \cite{koch16}. Our result thus holds for the most general conditions that can be imposed in the analysis of high-resolution quantizers.

The derivation of the lower bound reveals a necessary condition for a sequence of quantizers (parametrized by $D$) to achieve the asymptotic excess rate. Specifically, we demonstrated for scalar sources that the intersection of the quantization region $\set{S}_i$ with the interval $[\hat{x}_i-\rho D^{1/r},\hat{x}_i+\rho D^{1/r}]$ must have a Lebesgue measure that converges in probability to $2(1+r)^{1/r}D^{1/r}$ as $D\to 0$ and $\rho\to\infty$. This implies that any sequence of almost-regular quantizers achieving the asymptotic excess rate must converge in probability to a uniform quantizer as $D\to 0$. Since almost-regular quantizers achieve $D_{r,1}(R)$ when $r\geq 1$, this in turn suggests that asymptotically-optimal quantizers must essentially be uniform. 

While the presented bound is tight for scalar sources, it is unclear whether the same is true for multi-dimensional sources. Indeed, its derivation hinges on the fact that the distortion over the quantization region $\set{S}_i$ is lower-bounded by the distortion over a ball around $\hat{\vect{x}}_i$ with the same volume, cf.~\eqref{eq:proof_ULB}. Since the one-dimensional ball is an interval and, hence, tessellates $\Reals$, it follows that for one-dimensional sources the converse bound \eqref{eq:thm_main} is achieved by a tessellating quantizer (which in this case is the uniform quantizer). However, it is expected that this is no longer true for multi-dimensional sources, since in general balls do not tessellate the space. It is yet unclear whether there exists any (possibly non-tessellating) vector quantizer that achieves our converse bound for multi-dimensional sources.

\appendices

\section{Proof of Lemma~\ref{lemma:4}}
\label{app:lemma4}

To prove Lemma~\ref{lemma:4}, we first fix an arbitrary constant $\eta>0$ and divide the indices $i$ according to whether $\Lambda_{i,\eps} \geq \eta V_d \eps^d$ or not. Specifically, let
\begin{equation}
\set{I} \triangleq \left\{i\in\Integers\colon \Lambda_{i,\eps} \geq \eta V_d \eps^d\right\}
\end{equation}
and divide the sum on the LHS of \eqref{eq:proof_8} into
\begin{equation}
\label{eq:app_1}
\sum_{i} \Prob(\vect{X}\in\set{B}_{i,\eps}) \frac{\Lambda^{r/d}_{i,\eps}}{D} = \sum_{i\in\set{I}} \Prob(\vect{X}\in\set{B}_{i,\eps}) \frac{\Lambda^{r/d}_{i,\eps}}{D} + \sum_{i\in\cset{I}} \Prob(\vect{X}\in\set{B}_{i,\eps}) \frac{\Lambda^{r/d}_{i,\eps}}{D}
\end{equation}
where $\cset{I}$ denotes the complement of $\set{I}$. For every $i\in\cset{I}$ we have $\Lambda_{i,\eps}<\eta V_d\eps^d$, so the second sum on the RHS of \eqref{eq:app_1} can be upper-bounded as
\begin{IEEEeqnarray}{lCl}
\sum_{i\in\cset{I}} \Prob(\vect{X}\in\set{B}_{i,\eps}) \frac{\Lambda^{r/d}_{i,\eps}}{D} & \leq & \eta^{r/d} V_d^{r/d}\frac{\eps^r}{D} \sum_{i\in\cset{I}} \Prob(\vect{X}\in\set{B}_{i,\eps}) \nonumber\\
& \leq & \eta^{r/d} \frac{V_d^{r/d}}{\kappa} \label{eq:app_2}
\end{IEEEeqnarray}
where the second step follows because $\eps^r=D/\kappa$ and because, by definition, the sets $\set{B}_{i,\eps}$ are disjoint, so the sum of the probabilities $\Prob(\vect{X}\in\set{B}_{i,\eps})$ is equal to the probability of $\cup_{i\in\cset{I}}\set{B}_{i,\eps}$, which is upper-bounded by $1$.

To upper-bound the first sum on the RHS of \eqref{eq:app_1}, we begin by lower-bounding $\E{\|\vect{X}-\hat{\vect{X}}\|^r}$ as
\begin{IEEEeqnarray}{lCl}
\E{\|\vect{X}-\hat{\vect{X}}\|^r} & = & \sum_i \int_{\set{S}_i} f_{\vect{X}}(\vect{x}) \|\vect{x}-\hat{\vect{x}}_i\|^r \d \vect{x} \nonumber\\
& \geq & \sum_{i\in\set{I}} \int_{\set{B}_{i,\eps}} f_{\vect{X}}(\vect{x}) \|\vect{x}-\hat{\vect{x}}_i\|^r \d \vect{x} \nonumber\\
& = & \sum_{i\in\set{I}} \Prob(\vect{X}\in\set{B}_{i,\eps}) \frac{1}{\Lambda_{i,\eps}} \int_{\set{B}_{i,\eps}} \|\vect{x}-\hat{\vect{x}}_i\|^r \d \vect{x} \nonumber\\
& & {} - \sum_{i\in\set{I}} \int_{\set{B}_{i,\eps}}\left[\frac{1}{\Lambda_{i,\eps}}\Prob(\vect{X}\in\set{B}_{i,\eps})-f_{\vect{X}}(\vect{x})\right] \|\vect{x}-\hat{\vect{x}}_i\|^r \d \vect{x}.\label{eq:proof_8.5}
\end{IEEEeqnarray}
The region $\set{B}_{i,\eps}$ of volume $\Lambda_{i,\eps}$ that minimizes $\int_{\set{B}_{i,\eps}}\|\vect{x}-\hat{\vect{x}}\|^r\d \vect{x}$ is a ball around $\hat{\vect{x}}$. We thus have \cite[Section~III]{yamada80}
\begin{equation}
\label{eq:proof_ULB}
\frac{1}{\Lambda_{i,\eps}} \int_{\set{B}_{i,\eps}} \bigl\|\vect{x}-\hat{\vect{x}}_i\bigr\|^r \d \vect{x} \geq \frac{d}{d+r} \frac{\Lambda_{i,\eps}^{r/d}}{V_d^{r/d}}
\end{equation}
which yields for the first term on the RHS of \eqref{eq:proof_8.5}
\begin{equation}
\label{eq:proof_8.8}
 \sum_{i\in\set{I}} \Prob(\vect{X}\in\set{B}_{i,\eps}) \frac{1}{\Lambda_{i,\eps}} \int_{\set{B}_{i,\eps}} \|\vect{x}-\hat{\vect{x}}_i\|^r \d \vect{x} \geq \sum_{i\in\set{I}} \Prob(\vect{X}\in\set{B}_{i,\eps}) \frac{\Lambda_{i,\eps}^{r/d}}{V_d^{r/d}(1+r/d)}.
\end{equation}
Multiplying both sides of \eqref{eq:proof_8.5} by $V_d^{r/d}(1+r/d)/D$, applying \eqref{eq:proof_8.8} to \eqref{eq:proof_8.5}, and using that $\E{\|\vect{X}-\hat{\vect{X}}\|^r}\leq D$, we obtain
\begin{IEEEeqnarray}{lCl}
\sum_{i\in\set{I}} \Prob(\vect{X}\in\set{B}_{i,\eps}) \frac{\Lambda_{i,\eps}^{r/d}}{D} & \leq & V_d^{r/d}\left(1+\frac{r}{d}\right) \left(1+\frac{1}{D}\sum_{i\in\set{I}}  \int_{\set{B}_{i,\eps}} \left[\frac{1}{\Lambda_{i,\eps}}\Prob(\vect{X}\in\set{B}_{i,\eps})-f_{\vect{X}}(\vect{x})\right]\|\vect{x}-\hat{\vect{x}}_i\|^r \d \vect{x}\right). \IEEEeqnarraynumspace\label{eq:proof_bla}
\end{IEEEeqnarray}
We next introduce the pdf
\begin{IEEEeqnarray}{lCl}
f_{\vect{X}}^{(\Lambda)}(\vect{x}; \{\set{B}_{i,\eps}\}) & \triangleq & \sum_{i\in\set{I}} \frac{1}{\Lambda_{i,\eps}} \Prob(\vect{X}\in\set{B}_{i,\eps}) \I{\vect{x}\in\set{B}_{i,\eps}}\nonumber\\
& & {} +  f_{\vect{X}}(\vect{x})\left[\sum_{i\in\set{I}} \I{\vect{x}\in\bar{\set{B}}_{i,\eps}} + \sum_{i\in\cset{I}}\I{\vect{x}\in\set{S}_{i}}\right], \quad \vect{x}\in\Reals^d \IEEEeqnarraynumspace
\end{IEEEeqnarray}
which allows us to write
\begin{equation}
\sum_{i\in\set{I}}  \int_{\set{B}_{i,\eps}} \left[\frac{1}{\Lambda_{i,\eps}}\Prob(\vect{X}\in\set{B}_{i,\eps})-f_{\vect{X}}(\vect{x})\right]\|\vect{x}-\hat{\vect{x}}_i\|^r \d \vect{x} = \sum_i \int_{\set{S}_i} \left[f_{\vect{X}}^{(\Lambda)}(\vect{x}; \{\set{B}_{i,\eps}\})-f_{\vect{X}}(\vect{x})\right] \|\vect{x}-\hat{\vect{x}}_i\|^r \d \vect{x}.
\end{equation}
Since $\|\vect{x}-\hat{\vect{x}}_i\|^r\leq \eps^r=D/\kappa$ for $\vect{x}\in\set{B}_{i,\eps}$, $i\in\set{I}$ and $f_{\vect{X}}^{(\Lambda)}(\vect{x};\{\set{B}_{i,\eps}\})-f_{\vect{X}}(\vect{x})=0$ otherwise, we have
\begin{IEEEeqnarray}{lCl}
\left|\sum_i \int_{\set{S}_i} \left[f_{\vect{X}}^{(\Lambda)}(\vect{x};\{\set{B}_{i,\eps}\})-f_{\vect{X}}(\vect{x})\right] \|\vect{x}-\hat{\vect{x}}_i\|^r \d \vect{x}\right| & \leq & \frac{D}{\kappa} \int \left|f_{\vect{X}}^{(\Lambda)}(\vect{x};\{\set{B}_{i,\eps}\}) - f_{\vect{X}}(\vect{x})\right| \d \vect{x}.
\end{IEEEeqnarray}
Combining this upper bound with \eqref{eq:app_1}, \eqref{eq:app_2}, and \eqref{eq:proof_bla}, we obtain
\begin{equation}
\sum_{i} \Prob(\vect{X}\in\set{B}_{i,\eps}) \frac{\Lambda^{r/d}_{i,\eps}}{D} \leq V_d^{r/d}\left(1+\frac{r}{d}\right)\left(1+\frac{1}{\kappa}\int  \left|f_{\vect{X}}^{(\Lambda)}(\vect{x};\{\set{B}_{i,\eps}\}) - f_{\vect{X}}(\vect{x})\right| \d \vect{x} \right) + \eta^{r/d} \frac{V_d^{r/d}}{\kappa}. \label{eq:app_3}
\end{equation}
We next show that, for every $\eta>0$,
\begin{equation}
\label{eq:proof_really?}
\lim_{D\downarrow 0} \sup_{q(\cdot)} \int \left|f_{\vect{X}}^{(\Lambda)}(\vect{x};\{\set{B}_{i,\eps}\}) - f_{\vect{X}}(\vect{x})\right| \d \vect{x} = 0.
\end{equation}
(Note that $f_{\vect{X}}^{(\Lambda)}$ depends on $q(\cdot)$ and $D$ via $\set{B}_{i,\eps}$, $i\in\Integers$.) It then follows  that
\begin{equation}
\varlimsup_{D\downarrow 0} \sup_{q(\cdot)}\sum_{i} \Prob(\vect{X}\in\set{B}_{i,\eps}) \frac{\Lambda^{r/d}_{i,\eps}}{D} \leq V_d^{r/d}\left(1+\frac{r}{d}\right) + \eta^{r/d} \frac{V_d^{r/d}}{\kappa}
\end{equation}
which proves Lemma~\ref{lemma:4} upon letting $\eta$ tend to zero from above.

It thus remains to prove \eqref{eq:proof_really?}. By definition, $f_{\vect{X}}^{(\Lambda)}$ differs from $f_{\vect{X}}$ only when $\vect{x}\in\set{B}_{i,\eps}$, $i\in\set{I}$. Since the family of sets $\set{B}_{i,\eps}$, $i\in\set{I}$ (parametrized by $D$) has bounded eccentricity, it follows from Lebesgue's differentiation theorem that $f_{\vect{X}}^{(\Lambda)}$ converges to $f_{\vect{X}}$ almost everywhere as $D$ (and hence also $\eps$) tends to zero, which by Scheffe's lemma then implies \eqref{eq:proof_really?}. However, compared to the standard setting under which Lebesgue's differentiation theorem is proven, our setting is slightly more complicated, since as $D$ tends to zero not only the diameters of the sets $\set{B}_{i,\eps}$ decay, but also their locations in $\Reals^d$ may change. For completeness, we therefore provide all the steps, even though they follow closely the standard proof of the Lebesgue differentiation theorem.

We first note that the integral in \eqref{eq:proof_really?} is nonnegative and bounded, so its supremum is finite and for every $\nu>0$ there exists a sequence of quantizers (parametrized by $D$) such that
\begin{equation}
\label{eq:app_0.1}
\varlimsup_{D\downarrow 0}\int \left|f_{\vect{X}}^{(\Lambda)}(\vect{x};\{\set{B}_{i,\eps}\}) - f_{\vect{X}}(\vect{x})\right| \d \vect{x} \geq \varlimsup_{D\downarrow 0} \sup_{q(\cdot)} \int \left|f_{\vect{X}}^{(\Lambda)}(\vect{x};\{\set{B}_{i,\eps}\}) - f_{\vect{X}}(\vect{x})\right| \d \vect{x} - \nu.
\end{equation}
Since $\nu>0$ is arbitrary, it follows that, in order to prove \eqref{eq:proof_really?}, it suffices to show that for any sequence of quantizers (parametrized by $D$)
\begin{equation}
\label{eq:proof_yes!}
\lim_{D\downarrow 0}\int \left|f_{\vect{X}}^{(\Lambda)}(\vect{x};\{\set{B}_{i,\eps}\}) - f_{\vect{X}}(\vect{x})\right| \d \vect{x} = 0.
\end{equation}
Specifically, we shall show that for any sequence of quantizers (parametrized by $D$)
\begin{equation}
\label{eq:app_mu0}
\lambda\left(\left\{\vect{x}\in\Reals^d\colon \varlimsup_{D\downarrow 0} \left|f_{\vect{X}}^{(\Lambda)}(\vect{x};\{\set{B}_{i,\eps}\}) - f_{\vect{X}}(\vect{x})\right|>2\xi \right\}\right) = 0, \quad \text{for every $\xi>0$}
\end{equation}
where $\lambda(\cdot)$ denotes the Lebesgue measure on $\Reals^d$. It then follows that $f_{\vect{X}}^{(\Lambda)}$ converges to $f_{\vect{X}}$ almost everywhere as $D\to 0$ since
\begin{equation}
\left\{\vect{x}\in\Reals^d\colon \varlimsup_{D\downarrow 0} \left|f_{\vect{X}}^{(\Lambda)}(\vect{x};\{\set{B}_{i,\eps}\}) - f_{\vect{X}}(\vect{x})\right|>0 \right\} = \bigcup_{\ell=1}^\infty \left\{\vect{x}\in\Reals^d\colon \varlimsup_{D\downarrow 0} \left|f_{\vect{X}}^{(\Lambda)}(\vect{x};\{\set{B}_{i,\eps}\}) - f_{\vect{X}}(\vect{x})\right|> \frac{1}{\ell} \right\}
\end{equation}
and the countable union of sets of measure zero has measure zero. By Scheffe's lemma, almost everywhere convergence of $f_{\vect{X}}^{(\Lambda)}$ to $f_{\vect{X}}$ implies \eqref{eq:proof_yes!}, which together with \eqref{eq:app_0.1} proves the desired result \eqref{eq:proof_really?}.

We thus set out to prove \eqref{eq:app_mu0}. By the definition of $f_{\vect{X}}^{(\Lambda)}$ and the triangle inequality,
\begin{equation}
\label{eq:app_3.5}
\left| f_{\vect{X}}^{(\Lambda)}(\vect{x};\{\set{B}_{i,\eps}\}) - f_{\vect{X}}(\vect{x})\right| \leq \sum_{i\in\set{I}} \left|\frac{1}{\Lambda_{i,\eps}} \Prob(\vect{X}\in\set{B}_{i,\eps})-f_{\vect{X}}(\vect{x})\right| \I{\vect{x}\in\set{B}_{i,\eps}}.
\end{equation}
We next approximate $\frac{1}{\Lambda_{i,\eps}} \Prob(\vect{X}\in\set{B}_{i,\eps})$ by replacing $f_{\vect{X}}$ by a continuous function $g$. Indeed, since $f_{\vect{X}}$ is integrable, for every $\varepsilon>0$ there exists a continuous function $g$ such that \cite[Theorem~2.4.14, p.~92]{AsDo00}
\begin{equation}
\label{eq:app_L1dense}
\int |f_{\vect{X}}(\vect{x})-g(\vect{x})|\d\vect{x} \leq \varepsilon.
\end{equation}
It then follows that, for every $\vect{x}\in\set{B}_{i,\eps}$,
\begin{IEEEeqnarray}{lCl}
\left|\frac{1}{\Lambda_{i,\eps}} \Prob(\vect{X}\in\set{B}_{i,\eps}) - f_{\vect{X}}(\vect{x})\right| & \leq & \left|\frac{1}{\Lambda_{i,\eps}} \int_{\set{B}_{i,\eps}} g(\vect{y})\d\vect{y} - g(\vect{x})\right| \nonumber\\
& & {} +  \frac{1}{\Lambda_{i,\eps}}  \int_{\set{B}_{i,\eps}}\bigl|f_{\vect{X}}(\vect{y}) - g(\vect{y})\bigr|\d\vect{y} + \left|f_{\vect{X}}(\vect{x})-g(\vect{x})\right|. \IEEEeqnarraynumspace \label{eq:app_4}
\end{IEEEeqnarray}
Let $\set{B}(\vect{c},\rho)\triangleq \{\vect{x}\in\Reals^d\colon \|\vect{x}-\vect{c}\|\leq\rho\}$ denote the $d$-dimensional ball of radius $\rho$ centered at $\vect{c}$. Note that $\lambda\bigl(\set{B}(\hat{\vect{x}}_i,\eps)\bigr)= V_d \eps^d$. For every $\vect{x}\in\set{B}_{i,\eps}$ and $i\in\set{I}$, the second term on the RHS of \eqref{eq:app_4} can be upper-bounded by
\begin{IEEEeqnarray}{lCl}
\frac{1}{\Lambda_{i,\eps}}  \int_{\set{B}_{i,\eps}}\bigl|f_{\vect{X}}(\vect{y}) - g(\vect{y})\bigr|\d\vect{y} & \leq & \frac{1}{\eta\lambda\bigl(\set{B}(\hat{\vect{x}}_i,\eps)\bigr)} \int_{\set{B}(\hat{\vect{x}}_i,\eps)} \bigl|f_{\vect{X}}(\vect{y}) - g(\vect{y})\bigr|\d\vect{y} \nonumber\\
& \leq & \frac{2^d}{\eta} \frac{1}{\lambda\bigl(\set{B}(\vect{x},2\eps)\bigr)}\int_{\set{B}(\vect{x},2\eps)} \bigl|f_{\vect{X}}(\vect{y}) - g(\vect{y})\bigr|\d\vect{y} \nonumber\\
& \leq & \frac{2^d}{\eta} (f_{\vect{X}}-g)^{\star}(\vect{x}) \label{eq:app_5}
\end{IEEEeqnarray}
where $(f_{\vect{X}}-g)^{\star}$ denotes the Hardy-Littlewood maximal function for $f_{\vect{X}}-g$, i.e.,
\begin{equation}
(f_{\vect{X}}-g)^{\star}(\vect{x}) \triangleq \sup_{\rho>0}  \frac{1}{\lambda\bigl(\set{B}(\vect{x},\rho)\bigr)}\int_{\set{B}(\vect{x},\rho)} \bigl|f_{\vect{X}}(\vect{y}) - g(\vect{y})\bigr|\d\vect{y}, \quad \vect{x}\in\Reals^d.
\end{equation}
In \eqref{eq:app_5}, we have used that, for every $\vect{x}\in\set{B}_{i,\eps}$ and $i\in\set{I}$, we have  $\set{B}_{i,\eps}\subseteq \set{B}(\hat{\vect{x}}_i,\eps)\subseteq \set{B}(\vect{x},2\eps)$ and
\begin{equation*}
\Lambda_{i,\eps}\geq \eta \lambda\bigl(\set{B}(\hat{\vect{x}}_i,\eps)\bigr) = 2^{-d} \lambda\bigl(\set{B}(\vect{x},2\eps)\bigr).
\end{equation*}

Combining \eqref{eq:app_4} and \eqref{eq:app_5} with \eqref{eq:app_3.5}, we obtain
\begin{IEEEeqnarray}{lCl}
\IEEEeqnarraymulticol{3}{l}{\left| f_{\vect{X}}^{(\Lambda)}(\vect{x};\{\set{B}_{i,\eps}\}) - f_{\vect{X}}(\vect{x})\right|}\nonumber\\
\qquad  & \leq & \sum_{i\in\set{I}}  \left|\frac{1}{\Lambda_{i,\eps}} \int_{\set{B}_{i,\eps}} g(\vect{y})\d\vect{y} - g(\vect{x})\right|  \I{\vect{x}\in\set{B}_{i,\eps}}\nonumber\\
& & {} + \sum_{i\in\set{I}} \frac{2^d}{\eta} (f_{\vect{X}}-g)^{\star}(\vect{x})\I{\vect{x}\in\set{B}_{i,\eps}} + \sum_{i\in\set{I}} \left|f_{\vect{X}}(\vect{x})-g(\vect{x})\right| \I{\vect{x}\in\set{B}_{i,\eps}} \nonumber\\
& \leq & \sum_{i\in\set{I}}  \left|\frac{1}{\Lambda_{i,\eps}} \int_{\set{B}_{i,\eps}} g(\vect{y})\d\vect{y} - g(\vect{x})\right|  \I{\vect{x}\in\set{B}_{i,\eps}} + \frac{2^d}{\eta} (f_{\vect{X}}-g)^{\star}(\vect{x}) + \left|f_{\vect{X}}(\vect{x})-g(\vect{x})\right|, \quad \vect{x}\in\Reals^d \IEEEeqnarraynumspace \label{eq:app_6}
\end{IEEEeqnarray}
since the sets $\set{B}_{i,\eps}$, $i\in\set{I}$ are disjoint. The second and third term on the RHS of \eqref{eq:app_6} are independent of $D$ and $q(\cdot)$. The first term on the RHS of \eqref{eq:app_6} vanishes as $D$ tends to zero for any sequence of quantizers. Indeed, the continuity of $g$ implies that for every $\vartheta>0$ and $\vect{x}\in\Reals^d$ there exists an $\eps_0>0$ such that
\begin{equation}
|g(\vect{y})-g(\vect{x})| \leq \vartheta, \qquad \text{for $\|\vect{x}-\vect{y}\|\leq 2\eps_0$.}
\end{equation}
Since $\vect{x},\vect{y}\in\set{B}_{i,\eps}$ satisfy $\|\vect{x}-\vect{y}\|\leq 2\eps$, it follows that for every $\vartheta>0$ and $\vect{x}\in\Reals^d$ there exists an $\eps_0>0$ such that
\begin{equation}
\left|\frac{1}{\Lambda_{i,\eps}} \int_{\set{B}_{i,\eps}} g(\vect{y})\d\vect{y} - g(\vect{x})\right|  \I{\vect{x}\in\set{B}_{i,\eps}} \leq \vartheta \I{\vect{x}\in\set{B}_{i,\eps}}, \qquad \eps\leq \eps_0.
\end{equation}
Using that the sets $\set{B}_{i,\eps}$, $i\in\set{I}$ are disjoint, we conclude that for every $\vartheta>0$ and $\vect{x}\in\Reals^d$ there exists an $\eps_0>0$ such that
\begin{equation}
\sum_{i\in\set{I}}\left|\frac{1}{\Lambda_{i,\eps}} \int_{\set{B}_{i,\eps}} g(\vect{y})\d\vect{y} - g(\vect{x})\right|  \I{\vect{x}\in\set{B}_{i,\eps}} \leq \vartheta, \quad \eps\leq\eps_0.
\end{equation}
Since $\vartheta>0$ is arbitrary and $\eps$ vanishes as $D\to 0$, this implies that for every $\vect{x}\in\Reals^d$ and any sequence of quantizers
\begin{equation}
\label{eq:app_6.5}
\lim_{D\downarrow 0}\sum_{i\in\set{I}}  \left|\frac{1}{\Lambda_{i,\eps}} \int_{\set{B}_{i,\eps}} g(\vect{y})\d\vect{y} - g(\vect{x})\right|  \I{\vect{x}\in\set{B}_{i,\eps}} = 0.
\end{equation}

We conclude the proof of Lemma~\ref{lemma:4} by applying \eqref{eq:app_6} and \eqref{eq:app_6.5} to upper-bound the Lebesgue measure on the LHS of \eqref{eq:app_mu0}. Indeed, we have
\begin{IEEEeqnarray}{lCl}
\IEEEeqnarraymulticol{3}{l}{\lambda\left(\left\{\vect{x}\in\Reals^d\colon \varlimsup_{D\downarrow 0} \left|f_{\vect{X}}^{(\Lambda)}(\vect{x};\{\set{B}_{i,\eps}\}) - f_{\vect{X}}(\vect{x})\right|>2\xi \right\}\right)} \nonumber\\
\quad & \leq & \lambda\left(\left\{\vect{x}\in\Reals^d\colon \frac{2^d}{\eta} (f_{\vect{X}}-g)^{\star}(\vect{x}) + \left|f_{\vect{X}}(\vect{x})-g(\vect{x})\right| >2\xi \right\}\right) \nonumber\\
& \leq & \lambda\left(\left\{\vect{x}\in\Reals^d\colon \frac{2^d}{\eta} (f_{\vect{X}}-g)^{\star}(\vect{x}) >\xi \right\}\right) + \lambda\left(\left\{\vect{x}\in\Reals^d\colon \left|f_{\vect{X}}(\vect{x})-g(\vect{x})\right| >\xi \right\}\right).\label{eq:app_7}
\end{IEEEeqnarray}
The first term on the RHS of \eqref{eq:app_7} can be upper-bounded by using the Hardy-Littlewood maximal inequality \cite[Theorem~3.4, p.~55]{steinweiss71}
\begin{equation}
\label{eq:app_8}
\lambda\left(\left\{\vect{x}\in\Reals^d\colon \frac{2^d}{\eta} (f_{\vect{X}}-g)^{\star}(\vect{x}) >\xi \right\}\right) \leq \frac{2^d \alpha_d}{\eta\xi} \int |f_{\vect{X}}(\vect{x}) - g(\vect{x})|\d\vect{x}
\end{equation}
for some constant $\alpha_d$ that only depends on $d$. Likewise, the second term on the RHS of \eqref{eq:app_7} can be upper-bounded using Chebyshev's inequality \cite[Theorem~4.10.7, p.~192]{AsDo00}
\begin{equation}
\label{eq:app_9}
\lambda\left(\left\{\vect{x}\in\Reals^d\colon \left|f_{\vect{X}}(\vect{x})-g(\vect{x})\right| >\xi \right\}\right) \leq \frac{1}{\xi} \int |f_{\vect{X}}(\vect{x}) - g(\vect{x})|\d\vect{x}.
\end{equation}
Combining \eqref{eq:app_8} and \eqref{eq:app_9} with \eqref{eq:app_L1dense} and \eqref{eq:app_7}, it follows that
\begin{equation}
\label{eq:app_10}
\lambda\left(\left\{\vect{x}\in\Reals^d\colon \varlimsup_{D\downarrow 0} \left|f_{\vect{X}}^{(\Lambda)}(\vect{x};\{\set{B}_{i,\eps}\}) - f_{\vect{X}}(\vect{x})\right|>2\xi \right\}\right) \leq \frac{1+2^d\alpha_d/\eta}{\xi}\varepsilon.
\end{equation}
This proves \eqref{eq:app_mu0} upon letting $\varepsilon$ tend to zero from above, which was the last step required to prove Lemma~\ref{lemma:4}.

\section{Proof of Theorem~\ref{thm:necessary}}
\label{app:thmnecessary}
Following the steps \eqref{eq:mainproof_first}--\eqref{eq:proof_5} in the proof of Theorem~\ref{thm:main} in Section~\ref{sub:main_proof} particularized for $d=1$, we obtain that
\begin{IEEEeqnarray}{lCl}
H\bigl(q(X)\bigr)-R(D) & \geq & \frac{1}{r}\log\left(r2^r\Gamma(1+1/r)^r e\right) - \frac{1}{r}\sum_i \Prob(X\in\set{B}_{i,\eps}) \log\left(\frac{\Lambda_{i,\eps}}{D^{1/r}}+2\Gamma\left(\frac{1}{r},\frac{1}{\kappa\delta}\right)\right)^r \nonumber\\
 & & {} - \kappa \log\left(\frac{2}{\kappa^{1/r}}+2\Gamma(1/r)\right) - \kappa\left|\log\frac{r}{\delta^{1/r}}\right| - \frac{1}{\delta}. \label{eq:appB_1}
\end{IEEEeqnarray}
Recall that $\eps^r=D/\kappa$. The last three terms on the RHS of \eqref{eq:appB_1} are independent of $D$ and vanish as we first let $\kappa\to 0$ and then $\delta\to\infty$. To achieve
\begin{equation*}
\const{R}_{r,1} = \frac{1}{r} \log\left(\frac{\Gamma(1+1/r)^r e}{1+1/r}\right)
\end{equation*}
a sequence of quantizers (parametrized by $D$) must therefore satisfy
\begin{equation}
\label{eq:appB_2}
\varliminf_{\kappa\downarrow 0}\varlimsup_{D\downarrow 0}\frac{1}{r}\sum_i \Prob(X\in\set{B}_{i,\eps}) \log\left(\frac{\Lambda_{i,\eps}}{D^{1/r}}+2\Gamma\left(\frac{1}{r},\frac{1}{\kappa\delta}\right)\right)^r \geq \frac{1}{r} \log\bigl(2^r (1+r)\bigr).
\end{equation}
(As $\kappa\to 0$, the term on the LHS of \eqref{eq:appB_2} becomes independent of $\delta>0$.) For the sake of compactness, we shall use in the rest of the proof the following notation:\footnote{While all introduced quantities depend on $\kappa$, to keep the notation compact we only make the dependence on $D$ explicit.}

Let $V\triangleq 2^r(1+r)$. Further let $\upsilon \triangleq 2\Gamma\left(\frac{1}{r},\frac{1}{\kappa \delta}\right)$, and recall that $\lim_{\kappa\to 0}\upsilon=0$ for every $\delta>0$. Define
\begin{subequations}
\begin{IEEEeqnarray}{lCl}
\underline{\set{I}}_D & \triangleq & \left\{i\in\Integers\colon \frac{\Lambda_{i,\eps}^r}{D}\leq V-\vartheta\right\} \\
\overline{\set{I}}_D & \triangleq & \left\{i\in\Integers\colon \frac{\Lambda_{i,\eps}^r}{D}\geq V+\vartheta\right\}
\end{IEEEeqnarray}
\end{subequations}
and
\begin{subequations}
\begin{IEEEeqnarray}{lCllCl}
\underline{q}_D & \triangleq & \frac{1}{1-\wp_{\eps}} \sum_{i\in\underline{\set{I}}_D} \Prob(X\in\set{B}_{i,\eps}), \qquad &   \underline{\mu}_D & \triangleq & \frac{1}{(1-\wp_{\eps})\underline{q}_D} \sum_{i\in\underline{\set{I}}_D} \Prob(X\in\set{B}_{i,\eps}) \frac{\Lambda_{i,\eps}^r}{D}\\
\overline{q}_D & \triangleq & \frac{1}{1-\wp_{\eps}} \sum_{i\in\overline{\set{I}}_D} \Prob(X\in\set{B}_{i,\eps}), \qquad & \overline{\mu}_D & \triangleq & \frac{1}{(1-\wp_{\eps})\overline{q}_D} \sum_{i\in\overline{\set{I}}_D} \Prob(X\in\set{B}_{i,\eps}) \frac{\Lambda_{i,\eps}^r}{D} \\
\underline{\overline{q}}_D & \triangleq & \frac{1}{1-\wp_{\eps}} \sum_{i\in\Integers\setminus(\underline{\set{I}}_D\cup \overline{\set{I}}_D)} \Prob(X\in\set{B}_{i,\eps}), \qquad & \underline{\overline{\mu}}_D & \triangleq &  \frac{1}{(1-\wp_{\eps})\underline{\overline{q}}_D} \sum_{i\in\Integers\setminus(\underline{\set{I}}_D\cup \overline{\set{I}}_D)} \Prob(X\in\set{B}_{i,\eps}) \frac{\Lambda_{i,\eps}^r}{D}
\end{IEEEeqnarray}
\end{subequations}
where $\wp_{\eps}$ was defined in \eqref{eq:wpeps}. Finally, define
\begin{equation}
\mu_D \triangleq \underline{q}_D\,\underline{\mu}_D + \overline{q}_D\,\overline{\mu}_D + \underline{\overline{q}}_D\, \underline{\overline{\mu}}_D.
\end{equation}
By definition of $\underline{\set{I}}$ and $\overline{\set{I}}$, we have
\begin{equation}
\label{eq:appB_mubounds}
\underline{\mu}_D \leq V - \vartheta \qquad \text{and} \qquad \overline{\mu}_D \geq V+\vartheta.
\end{equation}
Furthermore, by Lemma~\ref{lemma:4} and \eqref{eq:proof_3},
\begin{equation}
\label{eq:appB_mu}
\varlimsup_{\kappa\downarrow 0}\varlimsup_{D\downarrow 0} \mu_D \leq V.
\end{equation}
Consequently, for any arbitrary $\varepsilon>0$, there exist $\kappa_0$ and $D_0$ such that
\begin{equation}
\label{eq:appB_3}
\mu_D \leq V + \varepsilon, \qquad (\kappa\leq\kappa_0,\, D\leq D_0).
\end{equation}
Without loss of generality, we implicitly assume that $\kappa$ and $D$ are sufficiently small, so that \eqref{eq:appB_3} holds.

We next apply steps similar to \eqref{eq:proof_6} and \eqref{eq:proof_6.5} to upper-bound
\begin{subequations}
\begin{IEEEeqnarray}{lCl}
\IEEEeqnarraymulticol{3}{l}{\frac{1}{1-\wp_{\eps}}\sum_i \Prob(X\in\set{B}_{i,\eps}) \log\left(\frac{\Lambda_{i,\eps}}{D^{1/r}}+\upsilon\right)^r} \nonumber\\
\qquad & \leq & \underline{q}_D \log\left(\underline{\mu}_D + \upsilon^r\right) + \overline{q}_D \log\Bigl(\overline{\mu}_D + \upsilon^r\Bigr) + \underline{\overline{q}}_D \log\left(\underline{\overline{\mu}}_D + \upsilon^r\right), \qquad \text{for $r<1$} \label{eq:appB_le1}
\end{IEEEeqnarray}
and
\begin{IEEEeqnarray}{lCl}
\IEEEeqnarraymulticol{3}{l}{\frac{1}{1-\wp_{\eps}}\sum_i \Prob(X\in\set{B}_{i,\eps}) \log\left(\frac{\Lambda_{i,\eps}}{D^{1/r}}+\upsilon\right)^r} \nonumber\\
\qquad & \leq & r\left[\underline{q}_D \log\left(\underline{\mu}_D^{1/r} + \upsilon\right) + \overline{q}_D \log\left(\overline{\mu}_D^{1/r} + \upsilon\right) + \underline{\overline{q}}_D \log\left(\underline{\overline{\mu}}_D^{1/r} + \upsilon\right)\right], \qquad \text{for $r\geq 1$.}\label{eq:appB_ge1}
\end{IEEEeqnarray}
\end{subequations}
It follows that, for $r<1$, any sequence of quantizers satisfying \eqref{eq:appB_2} must also satisfy
\begin{subequations}
\begin{IEEEeqnarray}{lCl}
\varliminf_{\kappa\downarrow 0} \varlimsup_{D\downarrow 0}\left\{\underline{q}_D \log\left(\underline{\mu}_D + \upsilon^r\right) + \overline{q}_D \log\left(\overline{\mu}_D + \upsilon^r\right) + \underline{\overline{q}}_D \log\left(\underline{\overline{\mu}}_D + \upsilon^r\right) - \log\left(V+\upsilon^r\right)\right\} \geq 0. \IEEEeqnarraynumspace \label{eq:appB_le2}
\end{IEEEeqnarray}
Likewise, for $r\geq 1$, any sequence of quantizers satisfying \eqref{eq:appB_2} must also satisfy
\begin{IEEEeqnarray}{lCl}
\varliminf_{\kappa\downarrow 0} \varlimsup_{D\downarrow 0}\left\{\underline{q}_D \log\left(\underline{\mu}_D^{1/r} + \upsilon\right) + \overline{q}_D \log\left(\overline{\mu}_D^{1/r} + \upsilon\right) + \underline{\overline{q}}_D \log\left(\underline{\overline{\mu}}_D^{1/r} + \upsilon\right) - \log\left(V^{1/r}+\upsilon\right)\right\} \geq 0. \IEEEeqnarraynumspace \label{eq:appB_ge2}
\end{IEEEeqnarray}
\end{subequations}

We conclude the proof of Theorem~\ref{thm:necessary} for the case $r\geq 1$ by demonstrating that any sequence of quantizers satisfying \eqref{eq:appB_ge2} must satisfy
\begin{equation}
\label{eq:appB_claim}
\lim_{\kappa\downarrow 0} \varlimsup_{D\downarrow 0} \underline{\overline{q}}_D = 1, \quad \text{for every $\vartheta>0$.}
\end{equation}
Substituting $\rho=1/\kappa$, this can be written as
\begin{equation}
\lim_{\rho\to\infty} \varlimsup_{D\downarrow 0} \frac{1}{1-\wp_{\eps}} \sum_{i} \Prob(X\in\set{B}_{i,\rho D^{1/r}}) \I{\left|\frac{\Lambda_{i,\rho D^{1/r}}^r}{D}- 2^r(1+r)\right|\leq\vartheta} = 1, \quad \text{for every $\vartheta>0$}
\end{equation}
which by Lemma~\ref{lemma:3} is equivalent to \eqref{eq:thm_necessary}. The proof for $r<1$ is almost identical and is therefore omitted.

To prove \eqref{eq:appB_claim} we use that, by the strict concavity of $x\mapsto \log x$, there exists a linear function $x\mapsto \ell_{x_0}(x)$ such that
\begin{equation}
\label{eq:appB_ell}
\log(x+\upsilon) \leq \ell_{x_0}(x), \quad x\geq 0
\end{equation}
with equality if, and only if, $x=x_0$. (Specifically, $\ell_{x_0}(x)=\frac{x+\upsilon}{x_0+\upsilon}+\log(x_0+\upsilon) -1$.) Moreover, we have
\begin{equation}
\log\left(\mu_D^{1/r}+\upsilon\right) \geq \underline{q}_D \ell_{\mu_D^{1/r}}\left(\underline{\mu}_D^{1/r}\right) + \overline{q}_D \ell_{\mu_D^{1/r}}\left(\overline{\mu}_D^{1/r}\right) + \underline{\overline{q}}_D \ell_{\mu_D^{1/r}}\left(\underline{\overline{\mu}}_D^{1/r}\right) \label{eq:appB_lin}
\end{equation}
since $x_0\mapsto \log(x_0+\upsilon) - \E{\ell_{x_0}(X)}$ (for any discrete random variable $X$) is monotonically increasing in $x_0$ and nonnegative for $x_0\geq \E{X}$, and since $\underline{q}_D\underline{\mu}_D^{1/r}+\overline{q}_D\overline{\mu}_D^{1/r} + \underline{\overline{q}}_D \underline{\overline{\mu}}_D^{1/r} \leq \mu_D^{1/r}$. The LHS of \eqref{eq:appB_ge2} can thus be upper-bounded by
\begin{multline}
\underline{q}_D\left[\log\left(\underline{\mu}_D^{1/r} + \upsilon\right)-\ell_{\mu_D^{1/r}}\left(\underline{\mu}_D^{1/r}\right)-\log\left(\frac{V^{1/r}+\upsilon}{\mu_D^{1/r}+\upsilon}\right)\right] \\
{} + \overline{q}_D\left[\log\left(\overline{\mu}_D^{1/r} + \upsilon\right)-\ell_{\mu_D^{1/r}}\left(\overline{\mu}_D^{1/r}\right)-\log\left(\frac{V^{1/r}+\upsilon}{\mu_D^{1/r}+\upsilon}\right)\right] \\
{} + \underline{\overline{q}}_D\left[\log\left(\underline{\overline{\mu}}_D^{1/r} + \upsilon\right)-\ell_{\mu_D^{1/r}}\left(\underline{\overline{\mu}}_D^{1/r}\right)-\log\left(\frac{V^{1/r}+\upsilon}{\mu_D^{1/r}+\upsilon}\right)\right]. \label{eq:appB_4}
\end{multline}
By \eqref{eq:appB_mu} and \eqref{eq:appB_ell}, the third term in \eqref{eq:appB_4} satisfies
\begin{equation}
\varlimsup_{\kappa\downarrow 0} \varlimsup_{D\downarrow 0} \underline{\overline{q}}_D\left[\log\left(\underline{\overline{\mu}}_D^{1/r} + \upsilon\right)-\ell_{\mu_D^{1/r}}\left(\underline{\overline{\mu}}_D^{1/r}\right)-\log\left(\frac{V^{1/r}+\upsilon}{\mu_D^{1/r}+\upsilon}\right)\right] \leq 0. \label{eq:appB_5}
\end{equation}
We further have
\begin{IEEEeqnarray}{lCl}
\IEEEeqnarraymulticol{3}{l}{\log\left(\underline{\mu}_D^{1/r} + \upsilon\right)-\ell_{\mu_D^{1/r}}\left(\underline{\mu}_D^{1/r}\right)-\log\left(\frac{V^{1/r}+\upsilon}{\mu_D^{1/r}+\upsilon}\right)} \nonumber \\
\qquad & \leq & \log\left((V-\vartheta)^{1/r}+\upsilon\right) - \ell_{(V+\varepsilon)^{1/r}}\left((V-\vartheta)^{1/r}\right) + \log\left(\frac{(V+\varepsilon)^{1/r} + \upsilon}{V^{1/r}+\upsilon}\right) \nonumber\\
& \triangleq & \underline{\mathsf{K}} \label{eq:appB_6}
\end{IEEEeqnarray}
and
\begin{IEEEeqnarray}{lCl}
\IEEEeqnarraymulticol{3}{l}{\log\left(\overline{\mu}_D^{1/r} + \upsilon\right)-\ell_{\mu_D^{1/r}}\left(\overline{\mu}_D^{1/r}\right)-\log\left(\frac{V^{1/r}+\upsilon}{\mu_D^{1/r}+\upsilon}\right)} \nonumber \\
\qquad & \leq & \log\left((V+\vartheta)^{1/r}+\upsilon\right) - \ell_{(V+\varepsilon)^{1/r}}\left((V+\vartheta)^{1/r}\right) + \log\left(\frac{(V+\varepsilon)^{1/r} + \upsilon}{V^{1/r}+\upsilon}\right) \nonumber\\
& \triangleq & \overline{\mathsf{K}}. \label{eq:appB_7}
\end{IEEEeqnarray}
Here, we used \eqref{eq:appB_mubounds} and \eqref{eq:appB_3} together with the facts that $x\mapsto \ell_{x_0}(x)-\log(x+\upsilon)$ is monotonically decreasing for $x\leq x_0$ and monotonically increasing for $x\geq x_0$, and $x_0\mapsto \log(x_0+\upsilon)-\ell_{x_0}(x)$ is monotonically increasing. 

Combining \eqref{eq:appB_4}--\eqref{eq:appB_7}, it follows that \eqref{eq:appB_ge2} can only be satisfied if
\begin{equation}
\varliminf_{\kappa\downarrow 0} \varlimsup_{D\downarrow 0} \max\left\{\underline{\mathsf{K}},\overline{\mathsf{K}}\right\}\left(\underline{q}_D+\overline{q}_D\right) \geq 0. \label{eq:appB_8}
\end{equation}
Since for $\varepsilon>0$ sufficiently small, we have
\begin{equation}
\lim_{\kappa\downarrow 0} \max\left\{\underline{\mathsf{K}},\overline{\mathsf{K}}\right\} < 0
\end{equation}
the condition \eqref{eq:appB_8}, in turn, can only be satisfied if
\begin{equation}
\lim_{\kappa\downarrow 0} \varliminf_{D\downarrow 0} \left(\underline{q}_D+\overline{q}_D\right) = 0.
\end{equation}
Using that $\underline{\overline{q}}_D=1-\underline{q}_D-\overline{q}_D$, the claim \eqref{eq:appB_claim} follows. This concludes the proof of Theorem~\ref{thm:necessary}.

\section*{Acknowledgment}
Stimulating discussions with Tam\'as Linder and Ram Zamir are gratefully acknowledged. The authors further wish to thank Giuseppe Durisi for calling their attention to reference \cite{wuverdu10_2}.

%\bibliographystyle{IEEEtran}           % in order of first citation
%\bibliography{header_short,bibliofile}


\end{document}